\documentclass[aps,prb,twocolumn,superscriptaddress,amsmath,amssymb,citeautoscript]{revtex4-2}

\usepackage{amsmath}
\usepackage{graphicx}
\usepackage[normalem]{ulem}
\usepackage{xcolor}
\usepackage{hyperref}
\hypersetup{colorlinks = {true},
    		citecolor  = {blue},
    		urlcolor   = {blue},
    		linkcolor  = {blue}}

\newcommand{\kph}{\ensuremath{\kappa_{\rm ph}}}
\newcommand{\sh}{\ensuremath{c_p}}
\newcommand{\Tmax}{\ensuremath{T_{\rm max}}}
\newcommand{\Tmin}{\ensuremath{T_1}}
\newcommand{\TSmax}{\ensuremath{T_2}}
\newcommand{\nH}{\ensuremath{n_{_{\rm H}}}}
\newcommand{\ryx}{\ensuremath{\rho_{yx}}}
\newcommand{\Sth}{\ensuremath{S_{\rm calc}}}
\newcommand{\Sexp}{\ensuremath{S_{\rm exp}}}
\newcommand{\bep}{\ensuremath{\beta^{\prime}}}

\newcommand{\Sup}{Supplemental Material (SM)}
\newcommand{\SI}{SM}

\newcommand{\ca}{Cd$_{3}$As$_{2}$}
\newcommand{\cza}{Cd$_{3-x}$Zn$_{x}$As$_{2}$}
\newcommand{\za}{Zn$_{3}$As$_{2}$}
\newcommand{\fom}{figure of merit}

\begin{document}

\title{Enhancement of the Thermoelectric Figure of Merit in the Dirac Semimetal \ca\ by Band-Structure and -Filling Control}

\author{Markus~Kriener}
\email[corresponding author: ]{markus.kriener@riken.jp}
\affiliation{RIKEN Center for Emergent Matter Science (CEMS), Wako 351-0198, Japan}
\author{Takashi~Koretsune}
\affiliation{Department of Physics, Tohoku University, Miyagi 980-8578, Japan}
\author{Ryotaro~Arita}
\affiliation{RIKEN Center for Emergent Matter Science (CEMS), Wako 351-0198, Japan}
\affiliation{Research Center for Advanced Science and Technology, University of Tokyo, Tokyo 153-8904, Japan}
\author{Yoshinori~Tokura}
\affiliation{RIKEN Center for Emergent Matter Science (CEMS), Wako 351-0198, Japan}
\affiliation{Department of Applied Physics and Quantum-Phase Electronics Center (QPEC), University of Tokyo, Tokyo 113-8656, Japan}
\affiliation{Tokyo College, University of Tokyo, Tokyo 113-8656, Japan}
\author{Yasujiro~Taguchi}
\affiliation{RIKEN Center for Emergent Matter Science (CEMS), Wako 351-0198, Japan}
\date{\today}

\begin{abstract}
Topological materials attract a considerable research interest because of their characteristic band structure giving rise to various new phenomena in quantum physics. Beside this, they are tempting from a functional materials point of view: Topological materials bear potential for an enhanced thermoelectric efficiency because they possess the required ingredients, such as intermediate carrier concentrations, large mobilities, heavy elements etc. Against this background, this work reports an enhanced thermoelectric performance of the topological Dirac semimetal \ca\ upon alloying the trivial semiconductor \za. This allows to gain fine-tuned control over both the band filling and the band topology in \cza. As a result, the thermoelectric \fom\ exceeds 0.5 around $x=0.6$ and $x=1.2$ at elevated temperatures. The former is due to an enhancement of the power factor, while the latter is a consequence of a strong suppression of the thermal conductivity. 
In addition, in terms of first-principle band structure calculations, the thermopower in this system is theoretically evaluated, which suggests that the topological aspects of the band structure change when traversing $x=1.2$.

\end{abstract}

\maketitle

\section{Introduction}
The development and exploration of sustainable energy devices is a timely task in materials design and research. One important issue is that a substantial amount of input energy is transferred into heat and lost irretrievably in many applications. To overcome this, the Seebeck effect offers a solution: It allows to regain electrical energy by conversion from waste heat. To quantify the conversion efficiency of bulk materials, the dimensionless \fom\ is used: $ZT = S^2 T/(\rho\kappa)$ with the thermopower $S$, the  resistivity $\rho$, and the  thermal conductivity $\kappa$ ($T$ denotes the temperature). The goal is to maximize $ZT$ \cite{rowe78a,mahan97a,snyder08a,ypei11a,heremans12a,jarman13a}. However, the involved quantities $\rho$, $S$, and $\kappa$ are strongly interrelated and difficult to tune independently, which explains why to date only a few materials with somewhat large $ZT$ values were found. Hence, there is a continuing demand on identifying new promising materials which ideally exhibit large $ZT$ values over a wide temperature range.

Various strategies and concepts to enhance $ZT$ have been proposed and employed, such as exploiting strong correlation effects \cite{terasaki97a,berggold05a}, carrier concentration control \cite{konstantinov01a,shimano17a,doi19a}, band structure engineering (band convergence and resonance level formation) \cite{ypei11a,heremans12a,ypei12a,qzhang13a,kriener15a,gtan15a,lwu17a,doi19a,khlee20a,fli24a,xshi24a,ygong24a}, magnetic interactions \cite{tsujii19a,vaney19a}, utilizing physical pressure \cite{nishimura19a}, or nanostructured devices \cite{poudel08a,kbiswas12a,qzhang13a}. ``Phonon glass + electron crystal'' is often referred to as the guiding principle, i.e., a material should possess ideally independent charge- and heat-transport channels to obtain small $\rho$ and $\kappa$ values simultaneously \cite{rowe95,snyder08a}.
\begin{figure*}[t]
\centering
\includegraphics[width=0.9\linewidth]{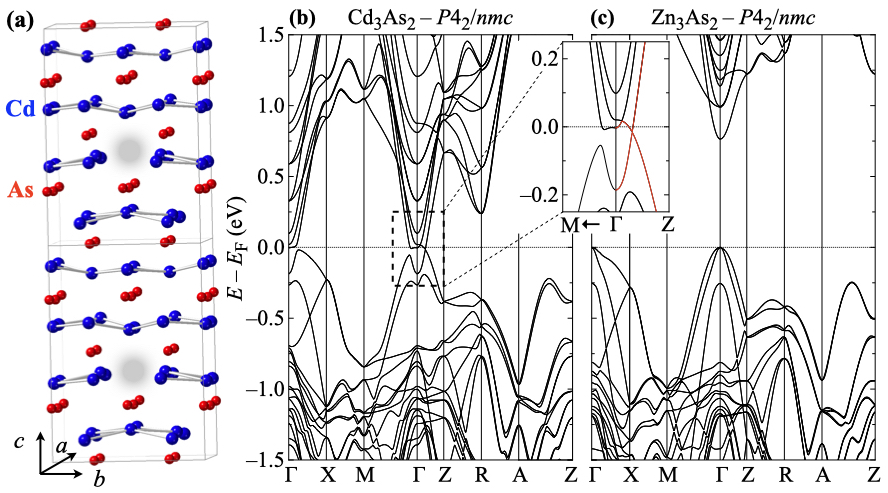}
\caption{(a) Schematic of \ca\ in its tetragonal $P4_2/nmc$ structure (``$\beta$-\ca''). In this setting, the characteristic voids (gray shadows) on the Cd sites arrange along the $a$ or $b$ axis, cf.\ Ref.~\cite{ali14a}. Band structures of (b) \ca\ and (c) \za\ based on the structural $\beta$ modification. In the inset, the band structure of \ca\ around the $\Gamma$ point is magnified with the Dirac-like band dispersion highlighted in red.}
\label{fig1}
\end{figure*}

Against this background, it is interesting that some materials with promising thermoelectric features have attracted attention from a different point of view in recent years: they were identified to possess a topologically nontrivial band structure, such as the prototypical thermoelectric Bi$_2$Te$_3$ \cite{muechler13a,ando15a,bansil16a}. The coincidence of topological nontriviality and thermoelectric efficiency is, however, not surprising since both phenomena partly share the same ingredients: intermediate carrier concentrations, large mobilities, heavy elements, or narrow band gaps \cite{takahashi10a,ghaemi10a}. This motivated us to focus on the Dirac semimetal \ca\  \cite{zwang13a,ali14a,tliang14a,zkliu14a,sjeon14a,uchida17a,nakazawa18a,uchida19a}. Beside its topological features, it has long been known for its large room-temperature mobility of $\sim 10^4 - 10^7$~cm$^2$V$^{-1}$s$^{-1}$ while the electron charge carrier concentration lies at moderate $\sim 10^{18}$cm$^{-3}$ \cite{turner61a,tliang14a}. Moreover, its thermal conductivity was reported to be as small as 2.5\,--\,4~WK$^{-1}$m$^{-1}$ at 300~K with an extremely tiny lattice contribution $\kph\sim 1$~WK$^{-1}$m$^{-1}$ \cite{spitzer66a,armitage69a,czhang16a,pariari18a,syue19a}. Consistently, a room-temperature \fom\ $ZT\sim 0.1$ was reported \cite{czhang16a,hwang19a}, with an additional slight enhancement upon shallow Cr doping \cite{czhang16a}. In thin films of \ca, enhanced values of the thermopower and the power factor $S^2/\rho$ were reported at low temperatures as a function of the film thickness \cite{wouyang24a}. There is also theoretical support that topological materials such as \ca\ are promising playgrounds to look for a reasonable thermoelectric performance \cite{hshi15a,tzhou16a,amarnath20a}.

Here, we chose to mix the $4d$ system \ca\ with its lighter $3d$ counterpart \za\ because partial Zn doping modifies both the band filling and the band structure: A hole carrier concentration of $\sim 10^{17}$~cm$^{-3}$ and a much smaller mobility of a few $10$~cm$^{2}$V$^{-1}$s$^{-1}$ reported for \za\ suggest that alloying offers to fine tune and gain control of these quantities and, hence, the \fom\ of \cza\ across the charge neutrality point \cite{zdanowicz64a,zdanowicz64b,zdanowicz75a,hlu17a}. In addition, the question at what $x$ the topological band inversion in \ca\ switches back to normal band order is tempting since \za\ is theoretically a trivial band insulator. Recent reports allocate this change around $x \sim 1.1$ for the bulk system \cite{hlu17a} while it was seen at a smaller $x\sim 0.6$ in thin-film studies \cite{nishihaya18a,nishihaya18b,nishihaya19a}.

In a preceding work \cite{fujioka21a}, where we only focussed on the thermoelectric performance of \cza\ below room temperature in the very limited Zn concentration range $0 \leq x \leq 1.2$, we reported that $ZT$ reaches $\sim 0.3\pm0.1$ for $x \sim 0.8 - 1.0$. This and the sizable slope of $ZT(T)$ at room temperature motivated the present work, in which we largely extend our previous study to approximately 650~K and target the whole solid solution $0 \leq x \leq 3$. Moreover, in the present work we also modeled the band structure on the basis of first-principle calculations for \ca\ and \za, and subsequently calculated the temperature dependence of the thermopower for various carrier concentrations for comparison with the experimental results to give an insight into the question where the topological  aspects of the band structure may change.

The examination of this much wider phase space in terms of temperature and Zn concentration allowed to identify two different $x$ ranges where $ZT$ becomes large: At 600~K, $ZT$ exceeds $0.5$ around $x\sim 0.6$ and $\sim 1.2$. In addition, the comparison of theoretical and experimental thermopower data yielded strong support for the conclusion that the topological band order vanishes between $x=1.0$ and 1.2. 
   
\section{Results}
\label{results}
We start with a brief discussion of the structure-related situation in \cza. At ambient conditions, a series of structural modifications is realized as a function of $x$. Moreover, for the end compounds \ca\ and \za,  two different modifications are discussed in the literature, referred to as $\alpha$ phase in this work: the body-centered tetragonal space groups $I4_1/acd$ (no.\ 142) \cite{zdanowicz64a,ali14a} and $I4_1/cd$ (no.\ 110) \cite{steigmann68a,pietraszko76a}. According to a recent publication, the former one seems realized in \ca\ \cite{ali14a}. In the Zn concentration range $0.6 \leq x \leq 1.2$, the system crystallizes in the simple tetragonal space group $P4_2/nmc$ (no.\ 137, $\beta$ phase) and for $1.5\leq x \leq 2.5$ it takes the closely related superstructure $I4_1/amd$ (no.\ 141; \bep\ phase). This complex situation is discussed in more detail in Section~S1 in the accompanying \Sup\ \cite{Suppl}. A schematic plot of the $\beta$-type crystal structure is shown in Fig.~\ref{fig1}(a).

Figures~\ref{fig1}(b) and (c) summarize the results of band-structure calculations for the two pristine materials \ca\ and \za, respectively. These are based on the $\beta$ phase which was chosen for an easy comparison with experimental data for alloyed samples. For \ca\ and \za, the main features of the $\beta$ phase are the same as in the $\alpha$ modifications. Notably, \ca\ remains a semimetal and exhibits a band crossing with Dirac-like dispersion near the $\Gamma$ point close to the Fermi energy as can be seen in the magnified inset in Fig.~\ref{fig1}. The opposite end compound \za\ is a trivial band insulator in both structural modifications, while \za\ is conductive due to unintentionally doped holes in reality. 

\begin{figure}[t]
\centering
\includegraphics[width=1\linewidth]{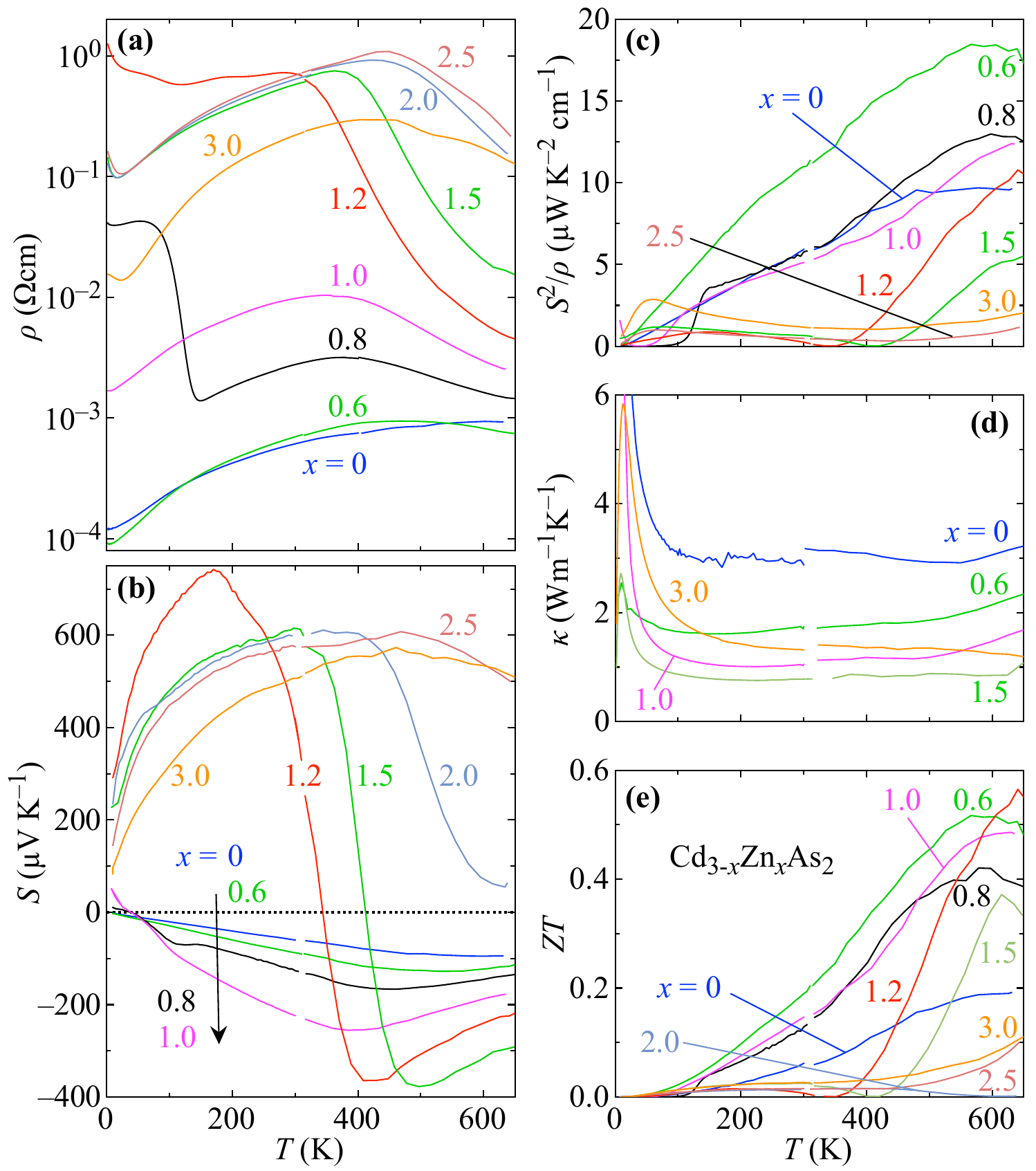}
\caption{Temperature dependence of the thermoelectric properties of the solid solution \cza\ ($0\leq x \leq 3$): (a) resistivity $\rho$, (b) thermopower $S$, (c) power factor $S^2/\rho$, (d) thermal conductivity $\kappa$, and (e) \fom\ $ZT = S^2 T/(\rho\kappa)$. In (c) the data for $x=2.0$ (overall very small $S^2/\rho$) and in (d) the data for some $x$ are omitted for clarity. The latter can be found in Section~S3 in the \SI\ \cite{Suppl}.}
\label{fig2}
\end{figure}
Figure~\ref{fig2} summarizes the temperature dependence of all quantities related to the \fom\ of \cza. The resistivity $\rho(T)$ is shown in Fig.~\ref{fig2}(a). Except for the strong increase for $T<\: \sim\! 150$~K in the data for $x=0.8$, all samples with $0.6\leq x\leq 1.0$ exhibit a metallic-like temperature dependence up to a temperature \Tmax\ and then decrease. As a function of $x$, \Tmax\ decreases monotonically, suggesting that there is also a maximum outside the targeted temperature range in $\rho$ for $x=0$, which tends to saturate above 600~K. As for the low-temperature increase in the data of $x=0.8$, we occasionally observe such metal-insulator transitions below room temperature for $x\leq 1.2$, cf.\ Ref.~\cite{fujioka21a} and Section~S7 in the \SI\ \cite{Suppl}.

For $x\geq 1.2$, the resistivity of \cza\ is strongly enhanced and exhibits maxima, which shift up to $x= 2.5$ toward higher temperatures. The data for $x\geq 1.5$ look qualitatively similar while the absolute values of the resistivity increases with $x$ except for \za, in rough agreement with the literature \cite{zdanowicz64a}. The maxima imply that additional charge carriers are activated at elevated temperatures, possibly leading to bipolar conduction. In Section~S5 in the \SI\ \cite{Suppl}, we present an Arrhenius analysis of the high-temperature resistivity. 

Figure~\ref{fig2}(b) summarizes corresponding thermopower $S(T)$ data. For all samples $x\leq 1$, we observe a negative $S$ except at very low temperatures. There are also distinct extrema whose temperatures \Tmin\ decrease monotonically with $x$. As expected for these metallic samples, $S$ exhibits roughly a linear temperature dependence below \Tmin\ with some deviations for $x=0.8$ and 1.0 at low temperatures. Above \Tmin, the absolute value of $S$ decreases. 
Upon further increasing $x$, there is a clear qualitative and quantitative change. For $x = 1.2$ and 1.5, $S$ exhibits two extrema at \Tmin\ and $\TSmax \;(< \Tmin)$. In between the sign of $S$ changes from positive to negative. For $T>\Tmin$, $S$ of these two samples is qualitatively similar to what is observed for $x\leq 1$. 
For $x=2.0$, there is no sign change of $S$ any more but the two extrema as characteristic features are still visible although $\Tmin(x)$ has almost shifted out of the examined temperature range. The latter seems to be the case for $x=2.5$ and 3, each of which exhibits only a broad extremum up to 650~K. 
The $x$ dependencies of these characteristic temperatures in $\rho(T)$ and $S(T)$ are replotted in Fig.~\ref{fig5}. 

Corresponding power-factor data $S^2/\rho(T)$ are plotted in Fig.~\ref{fig2}(c), which is strongly enhanced over the entire temperature range studied when going from $x=0$ to $0.6$. For $x = 0.8$ and 1.0, $S^2/\rho$ is similar to $x = 0$ up to above room temperature and exceeds it toward 650~K. 
For $x=1.2$ and 1.5, $S^2/\rho$ are much smaller than for $x\leq 1$ up to above room temperature, reflecting the large resistivities of these samples. Toward 650~K, a steep increase is observed for both and, in the case of $x=1.2$, it even exceeds $x=0$. 
For the sample with $x=2.0$, $S^2/\rho$ is rather small throughout the examined temperature range [not shown for clarity in Fig.~\ref{fig2}(c)]. For $x = 2.5$ and 3.0, there is a slight increase above $\sim 400$~K.

Figure~\ref{fig2}(d) summarizes thermal conductivity $\kappa(T)$ data. Beside the increases below $\sim 100$~K, $\kappa$ of \cza\ is remarkably small. For $x=0$, we observe $\kappa < 4$~WK$^{-1}$m$^{-1}$. Moreover, $\kappa$ is further suppressed when introducing Zn and takes its minimum for $x=1.5$ in most of the temperature range. Toward \za, $\kappa$ increases again but remains below that for $x=0$.

The \fom\ $ZT(T)$ for all examined samples is shown in Fig.~\ref{fig2}(e). For $0\leq x \leq 1.0$ we find relatively large $ZT$ values from well below room temperature, which further increase as a function of temperature up to about 500~K. Above, $ZT$ exhibits a maximum or an onset of a maximum. For $x = 1.2$ and 1.5, $ZT$ remains small up to $\sim 350$~K and 420~K, respectively. Upon further increasing temperature, these two samples also exhibit comparably large $ZT$ values. Interestingly, $ZT$ of $x=1.2$ eventually exceeds the data of $x=0.6$, although only above 600~K. For $x=2.5$ and 3.0, we find slightly enhanced $ZT$ values above $\sim 500$~K.

\section{Discussion}
\label{discussion}
\subsection{The Figure of Merit}
\begin{figure}[t]
\centering
\includegraphics[width=1\linewidth]{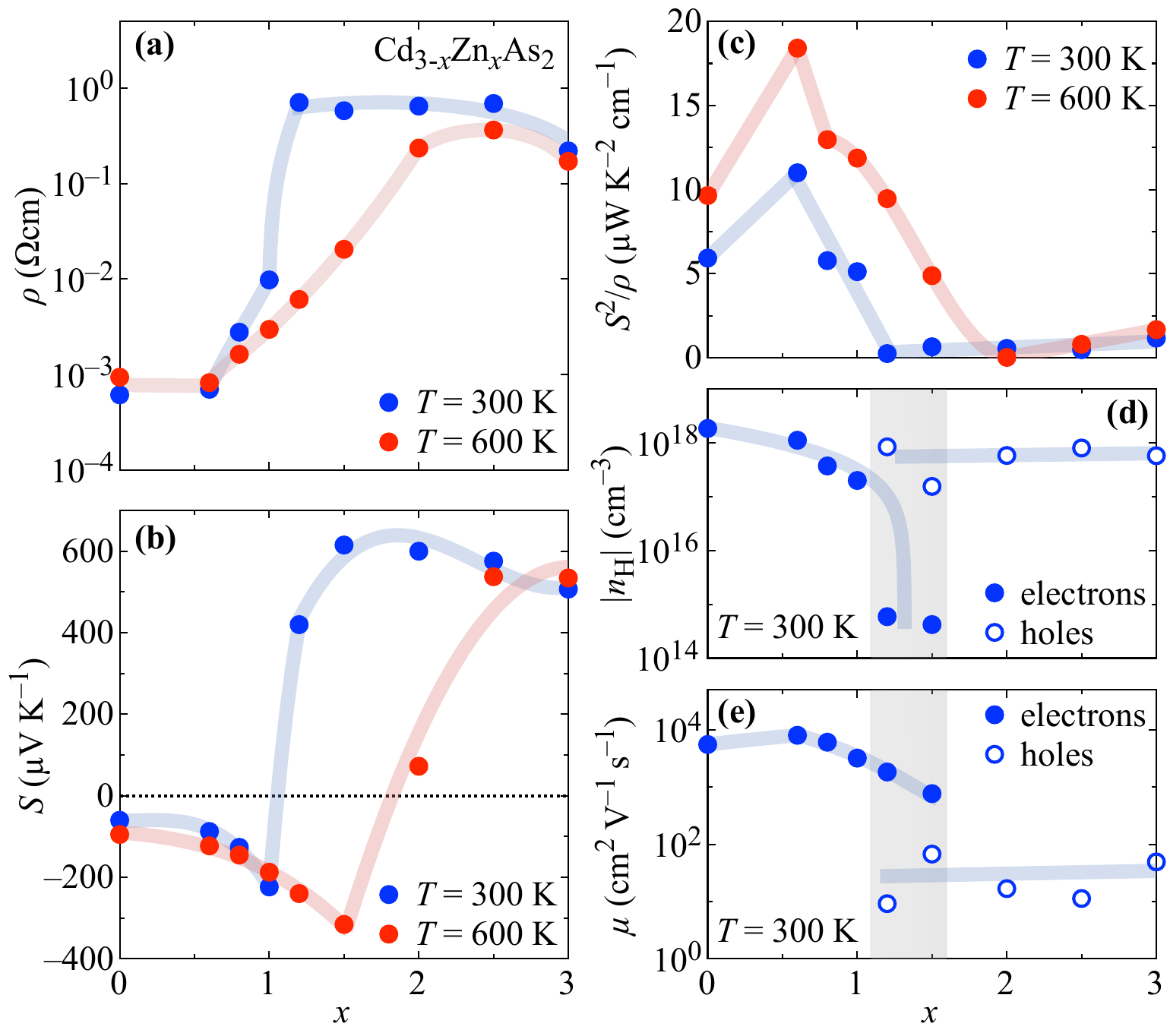}
\caption{Zn concentration $x$ dependence of (a) the resistivity $\rho$, (b) the thermopower $S$, (c) the power factor $S^2/\rho$, (d) the charge carrier concentration $|\nH|$ as estimated from field-dependent Hall-resistivity data, and (e) the corresponding charge carrier mobility $\mu$ of \cza\ ($0\leq x \leq 3$). Blue (red) symbols indicate data taken at 300~K (600~K). Blueish and reddish bold lines are guides to the eyes, and the gray shaded areas in (d) and (e) indicate the Zn concentration range in which electrons (filled symbols) and holes (open symbols) apparently coexist as charge carriers, see text.}
\label{fig3}
\end{figure}

\begin{figure}[t]
\centering
\includegraphics[width=1\linewidth]{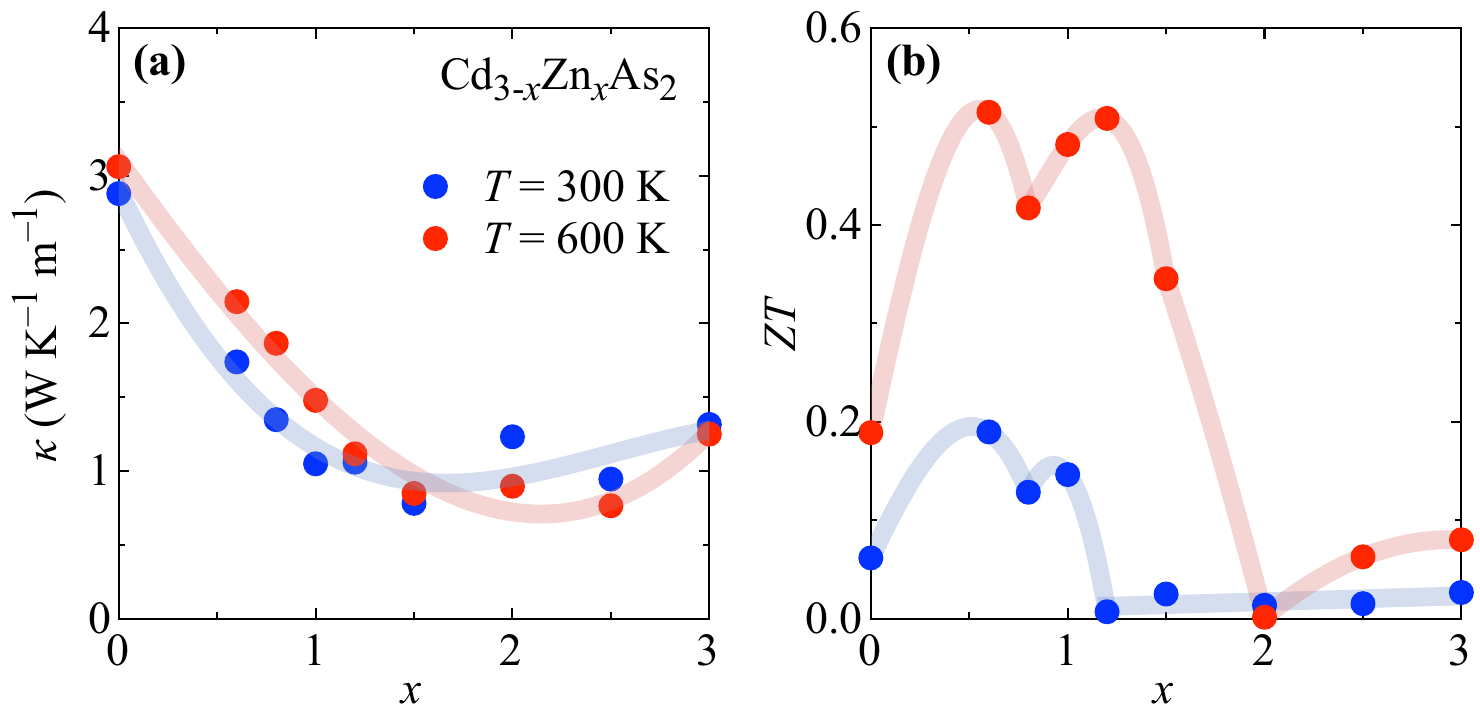}
\caption{Zn concentration $x$ dependence of (a) the thermal conductivity and (b) the dimensionless \fom\ $ZT$ of \cza\ ($0\leq x \leq 3$). Blue (red) symbols denote data taken at 300~K (600~K). Blueish and reddish bold lines are guides to the eyes.}
\label{fig4}
\end{figure}

Figures~\ref{fig3} and \ref{fig4} summarize several important quantities at 300~K and 600~K as a function of the Zn concentration $x$. In Fig.~\ref{fig3}(a), the strong enhancement of the resistivity $\rho(x)$ with $x$ is well recognizable in the 300-K data across $x=1$. In the 600-K data, this increase is less steep. A consequence of the maxima in $\rho(T)$ is that, except for $x \leq 0.6$, the resistivity values at 600~K are smaller than those at 300~K. 

Figure~\ref{fig3}(b) contains the respective thermopower data $S(x)$. The sign change of $S$ from negative ($x\leq 1.0$) to positive ($x\geq 2.0$) is clearly discernible. The occurrence of extrema above 300~K in $S(T)$ leads to comparable absolute values of the thermopower at 300~K and 600~K for $x\leq 1.0$ and $x\geq 2.5$. In between, the two extrema in $S(T)$ cause very different values of $S$ at 300~K and 600~K. For $x=1.5$, $S$ changes from about $+ 600$~\textmu VK$^{-1}$ to $-300$~\textmu VK$^{-1}$. 

These observations are reflected in the power-factor data $S^2/\rho(x)$ shown in Fig.~\ref{fig3}(c). The still relatively small resistivity and large absolute thermopower for $x\leq 1.0$ manifests itself as a strong enhancement of $S^2/\rho$ with a peak at $x=0.6$ at both temperatures shown. The recovery of the thermopower for $x \geq 2.5$ results only in a slight increase of $S^2/\rho$ toward \za\ due to the large resistivity. 

Figures~\ref{fig3}(d) and (e) summarize the $x$ dependencies of the absolute values of the charge carrier concentration $|\nH|$ as estimated from magnetic field $B$-dependent Hall-resistivity measurements $\ryx(B)$ at 300~K and the corresponding mobility $\mu=1/(|\nH| \rho e)$ with the elementary charge $e$, respectively. 

When introducing Zn into the system, the electronic transport remains initially dominated by electrons. Their concentration of $\sim 10^{18}$~cm$^{-3}$ for $x=0$ is reduced by approximately one order of magnitude for $x=1.0$, i.e., the band filling (with electrons as charge carriers) is reduced when introducing Zn into \ca.
Upon further alloying, the drop of the electron concentration is accelerated and falls below $10^{15}$~cm$^{-3}$ for $x=1.2$ and 1.5. At the same time, holes with a concentration of $10^{17}$\,--\,$10^{18}$~cm$^{-3}$ also start to contribute to the electronic transport, highlighting the change in the nature of the band filling as a function of $x$. The simultaneous presence of holes and electrons causes strongly non-linear Hall-resistivity data, cf.\ Figs.~S13 and S14 in Section~S4 in the \SI\ \cite{Suppl}. The temperature dependence of both carrier types for these two samples can be found therein in Fig.~S15. 
For $x\geq 2.0$, holes are the dominating charge carriers. Their concentration and, hence, the band filling does not change much. A simple phenomenological model of the band-filling evolution fitting to these observations is presented in Section~S6 in the \SI\ \cite{Suppl}. Another interesting point here is that the mobility of pristine \ca\ is apparently smaller than for $x=0.6$, above which it decreases again as a function of $x$. We will come back to this feature later.
\begin{figure*}[t]
\centering
\includegraphics[width=0.7\linewidth]{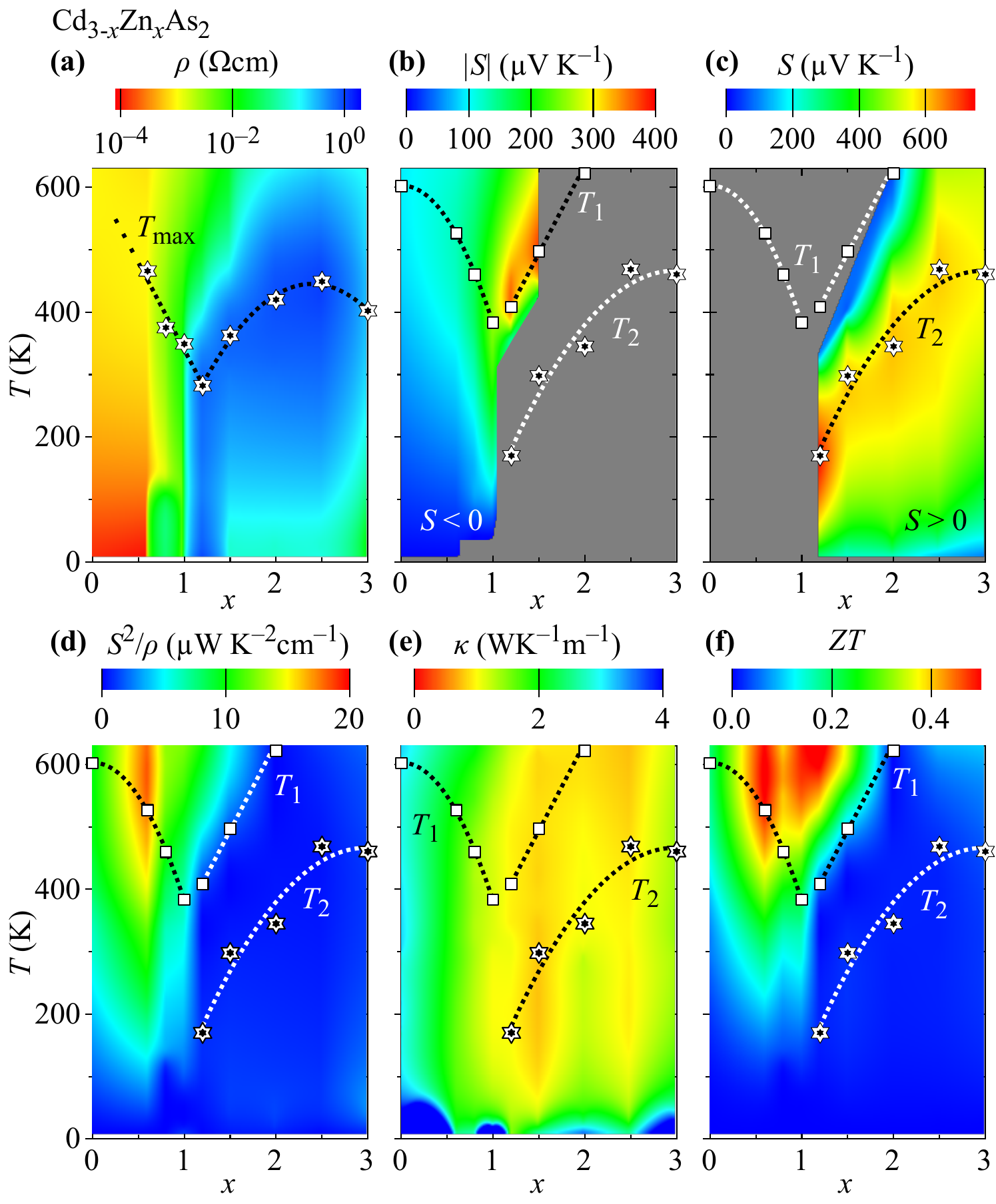}
\caption{Contour plots of all properties relevant to the \fom\ in \cza\ as a function of the Zn concentration $x$ (horizontal axis) and the temperature $T$ (vertical axis): (a) resistivity $\rho$, (b) and (c) absolute thermopower $|S|$ for $S<0$ and $S>0$, respectively (note the different scale bars), (d) power factor $S^2/\rho$, (e) thermal conductivity $\kappa$, and (f) \fom\ $ZT$. In panel (a), the maxima in the temperature dependence of the resistivity (\Tmax) and in panels (b)\,--\,(f) the respective extrema in temperature-dependent thermopower data (\Tmin\ and \TSmax) are superimposed. The color code in all panels is chosen so that red and blue indicate advantageous and disadvantageous for enhancing $ZT$, respectively.}
\label{fig5}
\end{figure*}

Importantly, the electrons governing the transport on the Cd-rich side of the phase diagram maintain their large mobility which amounts to several 1000~cm$^2$V$^{-1}$s$^{-1}$ for $x\leq 1.2$, cf.\ Fig.~\ref{fig3}(e). By contrast, the hole carriers for all samples $x\geq 1.2$ exhibit low mobility values  of less than 70~cm$^2$V$^{-1}$s$^{-1}$, i.e., Zn alloying removes high-mobility electrons and replaces them with low-mobility holes while the hole concentrations for $x\geq 1.2$ are on a similar order of magnitude as those of the electrons for $x\leq 1.0$. 

As seen in Fig.~\ref{fig4}(a), \cza\ exhibits an overall small $\kappa(x)$ at both 300~K and 600~K. The values at 300~K are smaller (larger) than those at 600~K for $x\leq 1.5$ ($x\geq 2.0$). The largest values of $\kappa \sim 3$~WK$^{-1}$m$^{-1}$ are found in pure \ca\ in agreement with previous reports in the literature \cite{spitzer66a,armitage69a,czhang16a,pariari18a,syue19a}.
Upon Zn alloying, $\kappa$ is further suppressed. We observe the smallest thermal conductivity $\kappa<1$~WK$^{-1}$m$^{-1}$ at 300~K for $x = 1.5$ and at 600~K for $x = 2.5$. This additional suppression of $\kappa$ is due to the vanishing electronic contributions for larger $x$, as it is discernible in Figs.~S10 and S12 in the \SI\ \cite{Suppl}.

Figure~\ref{fig4}(b) displays the \fom\ $ZT(x)$ of \cza. At 300~K, the enhancement of $S^2/\rho$ [Fig.~\ref{fig3}(c)] and the suppression of $\kappa$ [Fig.~\ref{fig4}(a)] lead to two peaks in $ZT(x)$. The first peak is observed at $x=0.6$, for which $ZT$ is close to 0.2. The second peak appears at $x=1.0$, where $ZT$ reaches $\sim 0.15$. For $x>1.0$, $ZT$ is almost negligible because of the large resistivity values measured for these samples around 300~K. 

At 600~K, The first peak is again observed at $x=0.6$ while the second peak has shifted to $x=1.2$. In both cases $ZT$ exceeds 0.5. The main driving ingredient in the former case is the power factor, while in the latter case the strong suppression of the thermal conductivity at intermediate $x$ provides the biggest impact. Hence, the drop of $ZT$ across $x=0.8$ at both temperatures is mainly due to the fading of the power factor in this $x$ range while the thermal conductivity has not yet been reduced sufficiently to compensate for it. For $x>1.2$, $ZT$ is strongly suppressed at 600~K, taking its lowest value at $x=2.0$. Upon further increasing $x$, a slight increase to almost $ZT \sim 0.1$ appears, which is mainly due to an again enhanced thermopower with $x$ at elevated temperatures in combination with the still fairly small thermal conductivity. 

Complementary contour plots are presented in Fig.~\ref{fig5}, which allow to easily perceive the overall evolution of the thermoelectric quantities and how they contribute to $ZT$ in \cza. The resistivity is shown in Fig.~\ref{fig5}(a). Its strong increase up to $x\sim\! 2.5$ at elevated temperatures and in the whole $x$ range at low temperatures are clearly revealed. 

Figures~\ref{fig5}(b) and (c) depict the absolute values of the thermopower $|S|$ for $S<0$ (smaller $x$) and $S>0$ (larger $x$), respectively. The initial increase of  $|S|$ with $x$ and $T$, when starting to introduce Zn, is traceable in Fig.~\ref{fig5}(b). The thermopower takes its largest absolute values at intermediate $x$ as indicated by the two red hot spots in the central upper half of Fig.~\ref{fig5}(b) and the central lower half of Fig.~\ref{fig5}(c), where the shift of \TSmax\ with $x$ toward higher temperatures accompanied by a slow reduction of $|S|$ is well resolved. The contour plot of the power factor $S^2/\rho$ shown in Fig.~\ref{fig5}(d) peaks around $x=0.6$ as functions of $x$ and $T$, making it apparent how $ZT(x)$ is enhanced at elevated temperatures for $x<\: \sim\!1.2$. 

The plot of the thermal conductivity in Fig.~\ref{fig5}(e) reveals how $\kappa$ is suppressed with $x$ and $T$ and that it remains small almost up to $x=3.0$, where some thermal conductivity is regained. The resulting \fom\ is shown in Fig.~\ref{fig5}(f), where the second peak in $ZT(x)$ around $x=1.2$ and at elevated temperatures is revealed and can be easily traced back to the concomitant suppression of $\kappa$.

\subsection{Topological Aspects of the Band Structure}
\begin{figure}[t]
\centering
\includegraphics[width=1\linewidth]{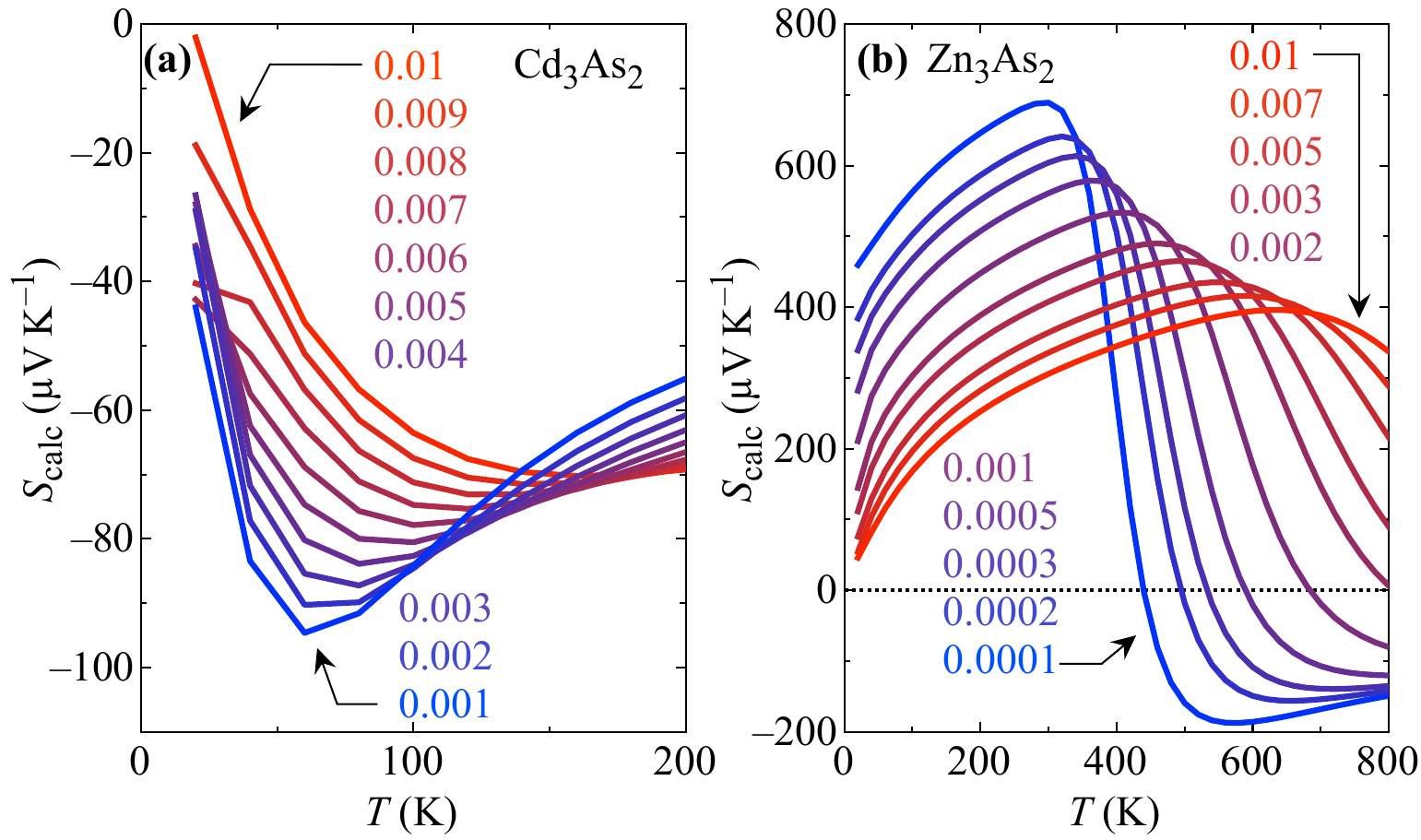}
\caption{Calculated temperature dependence of the thermopower \Sth\ of (a) \ca\ and (b) \za\ for different charge carrier densities [electrons in (a) and holes in (b), respectively] under the assumption of a rigid-band shift of the Fermi level. The change in the carrier count is indicated by the coloring (from large $\leftrightarrow$ red to small $\leftrightarrow$ blue). This reduction corresponds to the effect of the increase of the Zn (Cd) concentration in the experiment when starting from \ca\ (\za). The carrier counts are given in (a) in electrons per unit cell (corresponding to approximately $10^{18}$~cm$^{-3}\,-\,10^{19}$~cm$^{-3}$) and in (b) in holes per unit cell ($10^{17}$~cm$^{-3}\,-\,10^{19}$~cm$^{-3}$; 0.001 per unit cell corresponds to approximately $1 \times 10^{18}$cm$^{-3}$).}
\label{fig6}
\end{figure}

Finally, we discuss whether the anticipated changes in the band structure when alloying the Dirac semimetal \ca\ with the trivial semiconductor \za\ are reflected in our experimental observations shown in Figs.~\ref{fig2} -- \ref{fig4}. 
To this end, we theoretically calculated the thermopower as a function of temperature and for different charge carrier concentrations on the basis of the band structures for \ca\ and \za, presented in Figs.~\ref{fig1}(b) and (c), respectively. These assume the $\beta$ structural phase $P4_2/nmc$ which is realized for $0.6 \leq x \leq 1.2$. The \bep-phase, which forms for $1.5\leq x \leq 2.5$, has a superstructure in which the length of the crystallographic $c$ axis is doubled, cf.\ Section~S1 in the \SI\ \cite{Suppl}. The band structures as well as the band gaps are likely to be similar in both phases and, hence, also the resulting thermopower. Moreover, in the calculations to determine \Sth\ for \za, the band gap was reduced to 50\% of its obtained value (0.76 eV) by shifting the conduction bands in order to mimic the band structure for the intermediate region particularly around $1.2 \leq x \leq 1.5$, where the band gap is expected to be smaller than that of the end compound \za. It should also be noted that we apply the constant-$\tau$ (scattering time) approximation while the energy dependence of $\tau$ may contribute to $S$ significantly near the anticipated change in the band structure. 

Figures~\ref{fig6}(a) and \ref{fig6}(b) summarize the temperature dependencies of the calculated thermopower \Sth\ for several charge carrier concentrations for \ca\ and \za, respectively. The change in the carrier count in the calculations mimics the observed change in the experimental carrier concentration induced by alloying \ca\ and \za, as shown in Fig.~\ref{fig3}(d): Upon partially replacing Cd with Zn, the electron concentration is reduced while the hole concentration decreases upon introducing Cd into \za\ toward the intermediate $x$ range. 

A comparison of \Sth\ and the experimental results \Sexp\ shown in Fig.~\ref{fig2}(b) reveals striking similarities: In the case of \ca\ [Fig.~\ref{fig6}(a)], the theoretical curve with the largest electron concentration has to be compared with the experimental data for $x=0$: It is negative and exhibits an extremum as seen in the experimental thermopower in Fig.~\ref{fig2}(b). Upon reducing the carrier count, the absolute values $|\Sth|$ increase and the temperature of the extrema decreases, as seen in $\Sexp$ as a function of $x$ in Fig.~\ref{fig2}(b). Hence, there are qualitative and quantitative similarities between theory and experiment for $0\leq x \leq 1.0$ when the thermopower is calculated for \ca\ with a rigid-band assumption. 

As for the opposite end compound \za, the theoretical curve with the largest hole concentration in Fig.~\ref{fig6}(b) has to be compared with the experimental data for $x=3.0$, both of which exhibit a broad extremum. Upon reducing the carrier count, lower-$T$ extrema develop in $\Sth(T)$ and the temperatures of the higher-$T$ extrema decrease. Remarkably, the theoretical data for the smallest carrier counts almost perfectly reproduce the experimental results for $x = 1.2$ and 1.5 shown in Fig.~\ref{fig2}(b). Moreover, the calculations for \za\ do not only produce the order of magnitude of \Sexp\ but also the temperature scale fits well. Apparently, these theoretical data on \za\ fit qualitatively and quantitatively to the experimental results for $1.2\leq x \leq 3.0$. 

These observations provide hints that the physics in the solid solution \cza\ for $0\leq x \leq 1.0$ is likely governed by a band structure of the same type as realized in the Dirac semimetal \ca\ [Fig.~\ref{fig1}(b)], while for $1.2\leq x \leq 3.0$ it is of the same type as in \za\ [Fig.~\ref{fig1}(c)], i.e., the expected changes in the band structure take place between $x=1.0$ and 1.2.
Notably, several other properties also exhibit a significant change in this $x$ range. These are summarized and discussed in Section~S6 in the \SI\ \cite{Suppl} and therein shown in Fig.~S17. 

This finding also allows to speculate why the mobility $\mu$ shown in Fig.~\ref{fig3}(e) features a maximum at small Zn concentrations. The small effective mass $m^{*}$ for $x \sim 1.0$\,--\,1.2 with the chemical potential lying near the Dirac point should become larger when the chemical potential is raised upon increasing charge carrier concentration, i.e., decreasing $x$, cf. Fig.~\ref{fig3}(d). Hence, the mobility $\mu \propto \tau/m^{*}$ with the scattering time $\tau$ should decrease. However, experimentally the opposite is initially observed: $\mu$ increases upon decreasing $x$. This can be understood if the intrinsic disorder due to the mixture of Cd and Zn is taken into account (cf.\ the discussion in Section~S7 in the \SI): It is reasonable that the reduced disorder enhances $\tau$, overcompensating the expected effect of the effective mass enhancement upon decreasing $x$. When the disorder gets more and more reduced upon approaching \ca, at some $x$ this overcompensation will fade out and $\mu$ can be expected to exhibit a maximum, which we observe around $x \sim 0.6$.

We close the discussion with noting that such a strongly temperature-dependent thermopower including sign changes and the appearance of one or two extrema was also reported for Bi$_{1-x}$Sb$_{x}$ alloys, cf., e.g., \cite{ibrahim85a,lenoir98a}. This alloy is also bridging between semiconducting and semimetallic states and hosts the first experimentally confirmed three-dimensional topological insulator as a function of $x$ \cite{hsieh08a,ando13b}, showing a qualitative resemblance with the present \cza.

\section{Conclusion}
\label{summary}
To gain control over the thermoelectric \fom, this work follows the strategy to start with a topologically nontrivial system, here the Dirac semimetal \ca, and to alloy it with a trivial band insulator, namely its lighter counterpart \za. Both materials possess very different physical properties (e.g., metal vs.\ semiconductor, electron- vs.\ hole-type conduction, high vs.\ low mobility). Therefore, studying such a solid solution is a promising approach to gain fine-tuned control over the resistivity $\rho$, the thermopower $S$, the thermal conductivity $\kappa$, and, hence, the \fom\ $ZT = S^2 T/(\rho\kappa)$. The present work demonstrates that this approach has great potential to be successful: At elevated temperatures, the \fom\ of \cza\ exhibits two different $x$ ranges around $x=0.6$ and $1.2$, where $ZT$ exceeds $\sim 0.5$. While the first maximum in $ZT(x)$ can be traced back to an enhancement of the power factor, the second maximum is mainly caused by the suppression of the thermal conductivity. 

A comparison of experimental and theoretical thermopower data calculated as a function of temperature for various charge carrier concentrations points toward a scenario that the topological band structure present in the Dirac semimetal \ca\ changes to trivial between $x=1.0$ and 1.2. This is further supported by the observation that several other properties in this solid solution also exhibit significant changes in this $x$ range. This work underlines that topologically nontrivial systems are not only fascinating because of their intriguing physics, but that it can be also very promising and valuable to probe their thermoelectric performance especially when bridging toward trivial systems by alloying other elements.

\section{Methods}
\label{methods}
\subsection*{Sample growth and characterization}

Polycrystalline batches of \cza\ with $x = 0$, 0.6, 0.8, 1.0, 1.2, 1.5, 2.0, 2.5, and 3.0 were grown by conventional melt growth. For each batch, inside a glove box stoichiometric amounts of the respective elements (shots of Cd: purity 6N, Asahi Chemical Co.\ LTD., Japan; Zn: 5N, Nilaco Corp., Japan; As: 7N5, Furukawa Co.\ LTD., Japan) were loaded into carbonized quartz tubes. These were provisionally closed and transferred to a pumping station where they were evacuated and eventually sealed. Then they were composition-dependently fired to be melted for 48~h to 72~h at 950$^{\circ}$C to 1050$^{\circ}$C and eventually slowly cooled back to room temperature. The phase purity of all batches was checked with an in-house powder x-ray diffractometer (XRD, Rigaku, Japan). Our results are in qualitative agreement with the structural phase diagram reported in Ref.~\cite{zdanowicz64b}. A reproduction of this phase diagram is shown in Fig.~S1 in the \SI\ \cite{Suppl}. The concentrations $x$ examined in this work are highlighted therein. The results of the XRD measurements are shown in Figs.~S2 and S3 along with the lattice constants as a function of $x$ in Fig.~S4. Table~S1 provides an overview of the different structural phases existing in this solid solution and their relationship. 

The chemical composition of all batches was checked with a scanning-electron microscope equipped with an energy-dispersive x-ray analyzer (SEM-EDX, JEOL, Japan, and Bruker, USA). Resulting images for selected $x$ are shown in Figs.~S5\,--\,S7, and the results of the EDX analyses are summarized in Fig.~S8 in the \SI\ \cite{Suppl}. Throughout the text the nominal Zn concentration $x$ is used when referring to samples.

\subsection*{Measurements}
Various measurements at low and at elevated temperatures are necessary to determine the \fom. Some of them require different sample geometries. Therefore, for each $x$, several samples were cut from the respective as-grown batches.

\textit{Low-temperature measurements:} The measurements of the longitudinal $\rho$ and the Hall resistivity  $\rho_{yx}$ were performed by a conventional five-probe method in a commercial cryostat (physical property measurement system PPMS, Quantum Design, USA), employing the standard resistivity option. The analysis of the Hall resistivity is described in detail in Section~S4 in the \SI. The thermopower $S$ and the thermal conductivity $\kappa$ were measured in homebuilt setups fitting into a PPMS. A temperature gradient was applied to the long side of the respective sample and the voltage difference / temperature gradient was measured by commercial thermocouples, cf.\ Ref.~\cite{fujioka21a}.

\textit{High-temperature measurements:} Resistivity and thermopower were measured by employing a commercial apparatus (ZEM-3, Advance Riko, Japan) with the samples kept in He atmosphere and held by two Ni electrodes acting simultaneously as current leads. One of them is equipped with a heater to apply a temperature gradient. Two thermocouples acted as voltage pads in the resistivity measurements and were used to monitor the temperature gradient during the subsequent thermopower measurements. They were mechanically pushed onto a polished sample surface. The same samples used for the high-temperature $\rho$ and $S$ measurements were also used for the respective measurements at low temperatures, which were performed after the high-temperature experience.  

The thermal conductivity above room temperature was calculated via $\kappa= \lambda\,\sh\,d$ with the thermal diffusivity $\lambda$, the specific heat \sh, and the sample density at room temperature $d$ as estimated from the XRD data taken on each batch. To ensure that the XRD density is reliable, we roughly crosschecked it by also estimating the density of each sample from its mass and volume (from the linear dimensions). These two densities agree within $\pm 3$\%. The thermal diffusivity was measured in N$_2$ atmosphere by employing the laser-flash method in a commercial system (LFA-457, Netzsch, Germany). For this purpose, the polished top and bottom surfaces of the samples were coated with graphite spray. 
The specific heat was measured in Ar atmosphere in a commercial system by means of differential scanning calorimetry (STA449 F1 Jupiter, Netzsch, Germany).

The misfits between the low- and high-temperature data around room temperature are within 9.5\% (resistivity), 8.5\% (thermopower), and 15\% (thermal conductivity). 

We note that the upper limit of the temperature range studied here is due to the decomposition or degradation of \cza\ at elevated temperatures, as also reported earlier \cite{westmore64a,pietraszko73a}. The exact temperature range where the system dissociates differs between different studies. Our own analyses on powder samples suggest that a detectable mass loss sets in above 650~K. 

Comments concerning the known issues about the reproducibility of transport data in \ca\ (see, e.g., Refs.~\cite{tliang14a,crassee18a}) and the comparison of the present work with our preceding study \cite{fujioka21a} can be found in Section~S7 in the \SI\ \cite{Suppl}. 
We note for clarity that all batches were newly grown and that no data published in our previous study \cite{fujioka21a} are used in the present work while the examined temperature and Zn concentration ranges are largely extended.

\subsection*{Computational Details}
Electronic-structure calculations were performed in the space group $P4_2/nmc$ with the reported lattice constants for \ca\ \cite{pietraszko69a} and \za\ \cite{zdanowicz73a} by using the Vienna Ab initio Simulation Package (VASP) \cite{kresse96a}. For \ca, the projector-augmented wave (PAW) method \cite{bloechl94a,kresse99a} was employed with the exchange correlation functionals of the Perdew-Burke-Ernzerhof (PBE) type \cite{perdew96a} in agreement with a previous calculation \cite{zwang13a} and for \za\ the modified Becke-Johnson (mBJ) type \cite{becke06a,tran09a}, respectively. By using mBJ, we obtain a band gap of 0.76~eV for \za, which is basically consistent with previous calculations \cite{ullah17a,hnuna23a}. Spin-orbit coupling is included in all the calculations using VASP. We utilized the BoltzWann module \cite{pizzi14a} of the Wannier90 package \cite{pizzi20a} to calculate the thermopower using a $k$-point mesh of $100 \times 100 \times 100$ and employ the constant-$\tau$ approximation, i.e., the energy dependence of the scattering rate is neglected for simplicity.

\section*{Acknowledgement}
MK acknowledges fruitful discussions with and support by H.~Kobayashi (Netzsch Japan), J.~Hanss (Netzsch Germany), and D.~Maryenko. 
This work was partly supported by the Japan Society for the Promotion of Science (JSPS, No.\ 24224009, 15K05140, 19H05825, 22K03447, 22K18954, and 21H01003). This work was also supported by the RIKEN TRIP initiative (Many-body Electron Systems).

\end{document}


\title{Supplemental Material for \\ ``Enhancement of the Thermoelectric Figure of Merit in the Dirac Semimetal \ca\ by Band-Structure and -Filling Control''}

\author{Markus~Kriener}
\email[corresponding author: ]{markus.kriener@riken.jp}
\affiliation{RIKEN Center for Emergent Matter Science (CEMS), Wako 351-0198, Japan}
\author{Takashi~Koretsune}
\affiliation{Department of Physics, Tohoku University, Miyagi 980-8578, Japan}
\author{Ryotaro~Arita}
\affiliation{RIKEN Center for Emergent Matter Science (CEMS), Wako 351-0198, Japan}
\affiliation{Research Center for Advanced Science and Technology, University of Tokyo, Tokyo 153-8904, Japan}
\author{Yoshinori~Tokura}
\affiliation{RIKEN Center for Emergent Matter Science (CEMS), Wako 351-0198, Japan}
\affiliation{Department of Applied Physics and Quantum-Phase Electronics Center (QPEC), University of Tokyo, Tokyo 113-8656, Japan}
\affiliation{Tokyo College, University of Tokyo, Tokyo 113-8656, Japan}
\author{Yasujiro~Taguchi}
\affiliation{RIKEN Center for Emergent Matter Science (CEMS), Wako 351-0198, Japan}
\date{\today}

\maketitle

This \Sup\ provides additional data on \cza\ as follows:

\begin{itemize}
\item Section~\ref{xrd}: About the crystal structures in \cza
\item Section~\ref{edx}: Results of SEM-EDX analyses
\item Section~\ref{kap}: Thermal conductivity data
\item Section~\ref{rhoyx}: Analysis of the Hall resistivity for $x=1.2$ and 1.5
\item Section~\ref{arrh}: Analysis of the longitudinal resistivity at high temperatures
\item Section~\ref{comp}: Comparison of additional properties
\item Section~\ref{reprod}: Sample dependence of transport data in \cza
\end{itemize}

\clearpage

\subsection{About the crystal structures in \cza}
\label{xrd}
\subsubsection*{\textbf{Structural phase diagram from Ref.~\cite{zdanowicz64b}}}
\begin{figure}[h!]
\centering
\includegraphics[width=0.8\linewidth,clip]{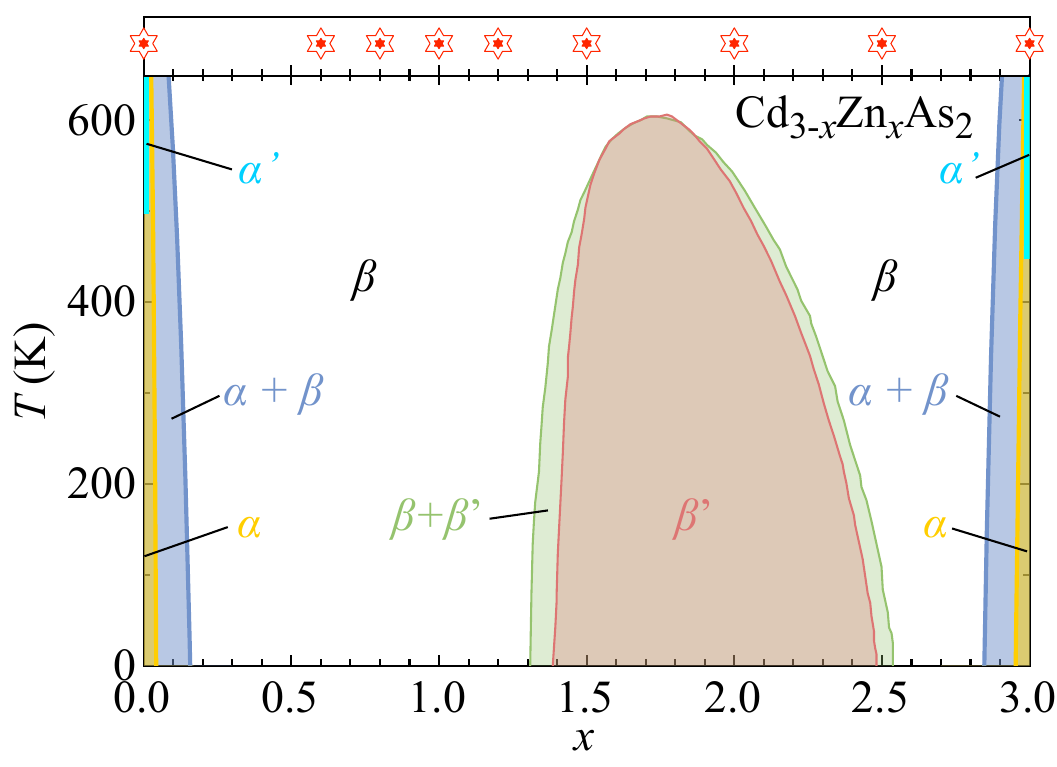}
\caption{Structural phase diagram replotted from Ref.~\cite{zdanowicz64b}. For the definition of the different phases see text. The red star symbols at the top indicate the nine different Zn concentrations studied in this work.}
\label{figS1}
\end{figure}
Figure~\ref{figS1} is replotted from Ref.~\cite{zdanowicz64b} and provides an overview of the complex situation on the crystal structures realized in \cza. In addition, the Zn concentrations of the samples studied in this work are indicated by red star symbols at the top of the diagram.
\begin{table}[t]
\centering
\caption{\textbf{Different crystal structures realized in Cd\boldmath{$_{3-x}$}Zn\boldmath{$_{x}$}As\boldmath{$_{2}$.}}}
\label{tabS1}
\begin{tabular}{ccccccc}
\toprule
Label                   & Crystal    & Space  & \multicolumn{2}{c}{Lattice}   & $Z$ & Reference                         \\ \addlinespace[-0.75em]
                        & Structure  & Group  & \multicolumn{2}{c}{Parameter} &     &                                   \\ \hline \addlinespace[0.75em]
\multirow{2}*{$\alpha$} & $I4_1/acd$ & 142    & $a$            & $c$          & 32  & \cite{zdanowicz64b,ali14a}        \\ \addlinespace[0.1em]
                        & $I4_1/cd$  & 110    & $a$            & $c$          & 32  & \cite{steigmann68a,pietraszko76a} \\ \addlinespace[0.5em]
$\alpha^{\prime}$       & $P4_2/nbc$ & 133    & $\approx a$    & $\approx c$  & 32  & \cite{pietraszko73a}              \\ \addlinespace[0.5em]
$\beta$                 & $P4_2/nmc$ & 137    & $a/\sqrt{2}$ & $c/2$          &  8  & \cite{zdanowicz64b,zdanowicz73a}  \\ \addlinespace[0.5em]
\bep\                   & $I4_1/amd$ & 141    & $a/\sqrt{2}$ & $c$            & 16  & \cite{zdanowicz64b,volodina13a}   \\ \addlinespace[0.5em]
\bottomrule\addlinespace[0.5em]
\end{tabular}
\end{table}

\cza\ exhibits structural phase transitions as a function of $x$ and temperature. The different structures relevant here are all tetragonal and summarized in Table~\ref{tabS1}. Therein, the tetragonal lattice constants $a$ and $c$ are given in relation to the structural modification labeled $\alpha$, and $Z$ denotes the number of formula units per unit cell. Note that the labels for these structural modifications are not used consistently in the literature. The present notation follows Ref.~\cite{zdanowicz64b}. In addition to these listed in Table~\ref{tabS1}, for \ca\ there are also two additional high-temperature modifications ($T>650$~K) mentioned in the literature: $\alpha^{\prime\prime}$ (tetragonal) and a cubic one. For \za, only the cubic phase is reported \cite{zdanowicz64b,pietraszko73a,ali14a}. According to our own measurements, the system starts to dissociate above approximately 650~K, for which reason these are omitted here. The dissociation process is also mentioned in \cite{westmore64a,pietraszko73a}. 

\subsubsection*{\textbf{X-ray diffraction patterns}}
\begin{figure}[h!]
\centering
\includegraphics[width=0.85\linewidth,clip]{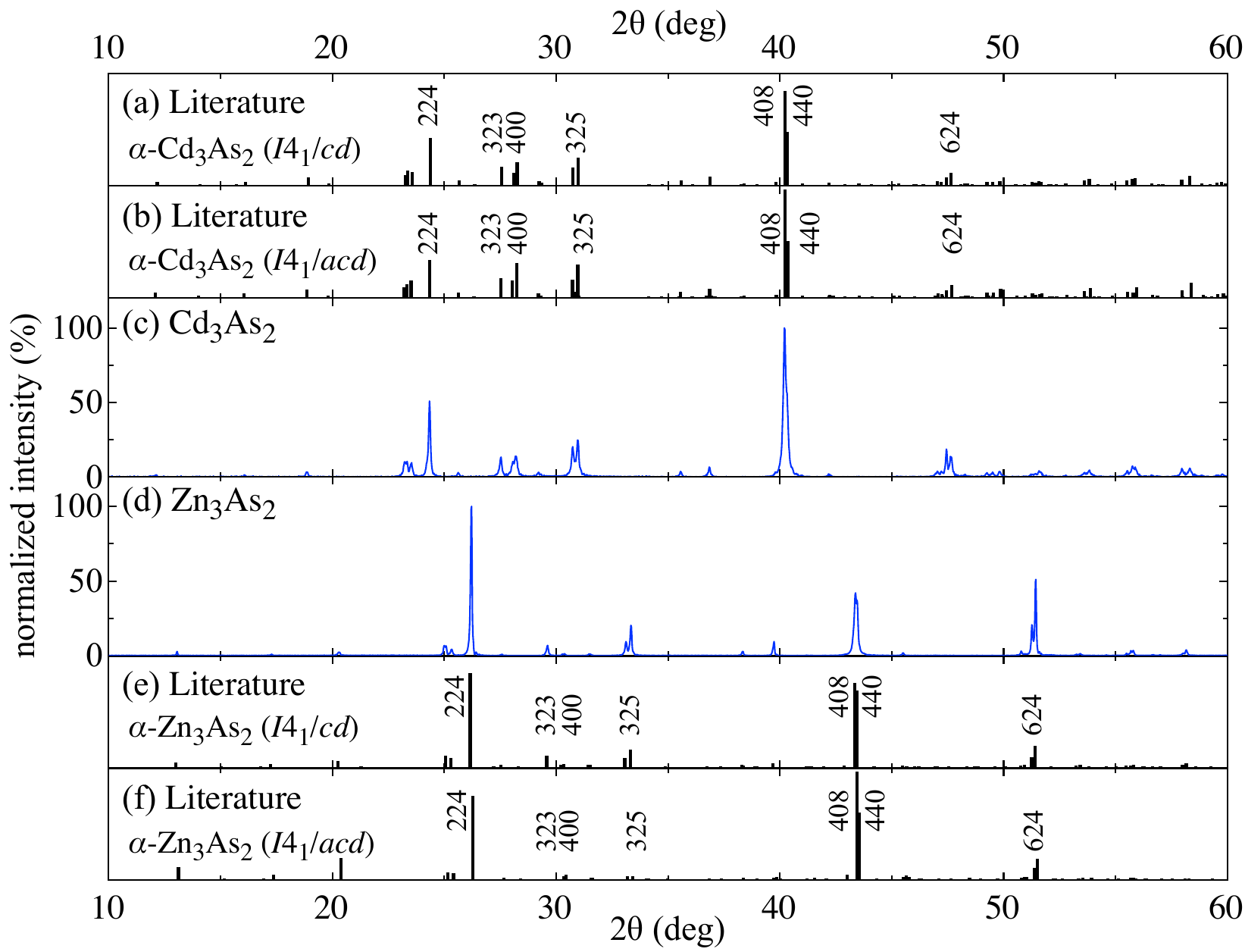}
\caption{Normalized XRD data of the two pristine compounds (c) \ca\ and (d) \za. The expected peak positions for both structures discussed in the literature are shown for \ca\ in (a) $I4_1/cd$ and (b) $I4_1/acd$ and those for \za\ in (e) and (f), respectively. Stronger reflections are labeled.}
\label{figS2}
\end{figure}

\begin{figure}[h!]
\centering
\includegraphics[width=0.85\linewidth,clip]{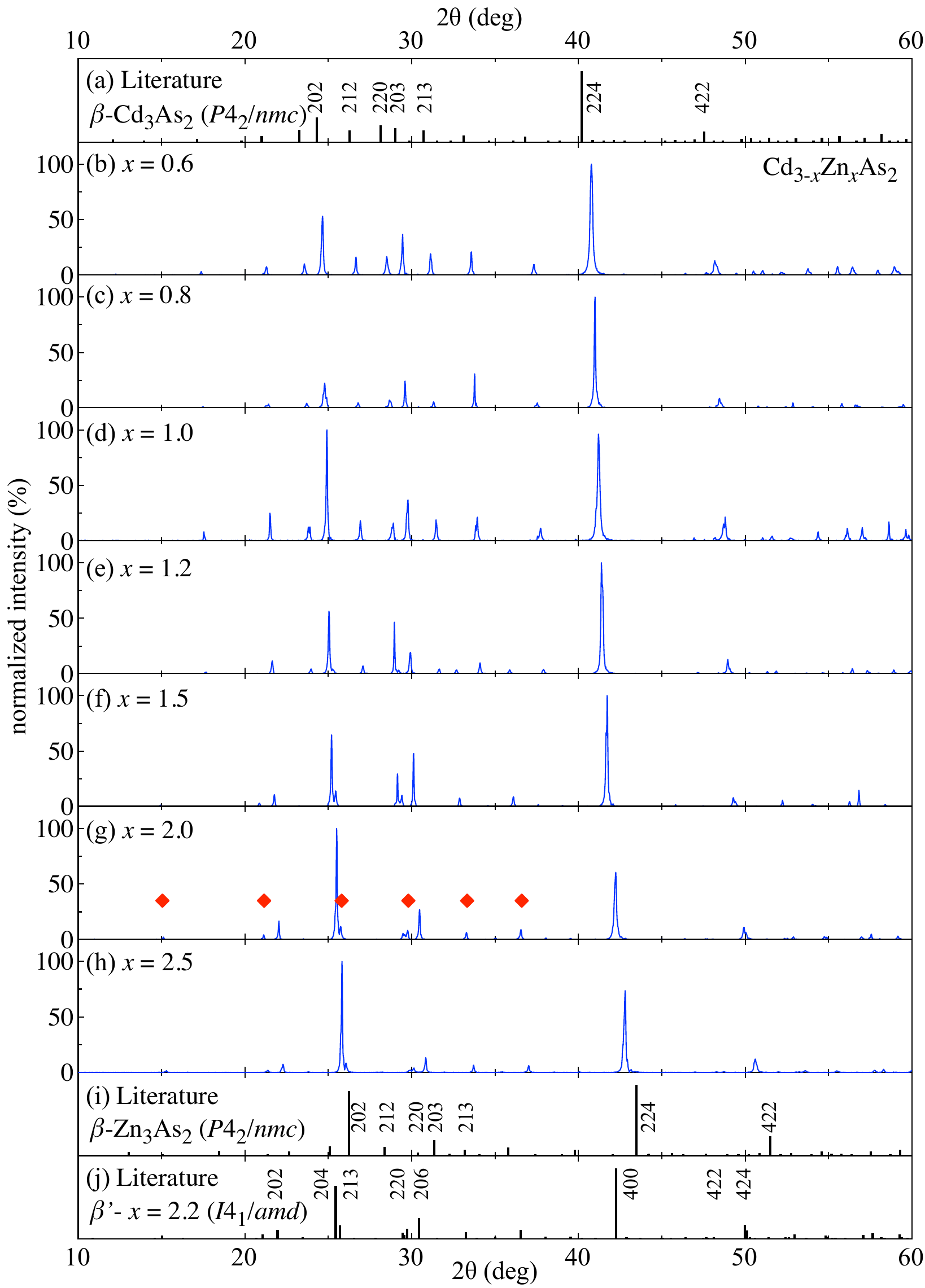}
\caption{Normalized XRD data of the seven alloyed compounds \cza\ with $0.6 \leq x \leq 2.5$ studied here. The expected peak positions for $\beta$-\ca\ and $\beta$-\za\ are shown in (a) and (i), respectively. Panel (j) shows the peak positions of a sample with $x=2.2$ in the \bep\ structure from Ref.~\cite{volodina13a} for comparison. In panel (g), some of the peaks belonging to \bep\ are indicated with red diamonds. Stronger reflections are labeled. See text for details.}
\label{figS3}
\end{figure}

In this subsection the x-ray diffraction (XRD) patterns taken at room temperature of all nine studied batches are discussed. Since both pristine compounds \ca\ and \za\ take a different structure as compared to the alloyed crystals, the XRD patterns are split into Figures~\ref{figS2} and \ref{figS3}.
Figure~\ref{figS2} presents XRD data of \ca\ in panel (c) and \za\ in (d) normalized to the respective main peaks. The expected peak positions for the two settings discussed in the literature are shown for \ca\ in (a) $I4_1/cd$ and (b) $I4_1/acd$ and for \za\ in (e) and (f), respectively.
\clearpage
Figure~\ref{figS3} summarizes XRD data for the seven alloyed batches of \cza\ ($0.6\leq x \leq 2.5$) normalized to the respective main peaks.  Panels (a) and (i) show the expected peak positions for \ca\ and \za\ in the $\beta$ phase $P4_2/nmc$, (j) provides the expected peak positions for the \bep\ structure $I4_1/amd$ for a sample with $x=2.2$ taken from a recent publication \cite{volodina13a} which can be readily compared with the data in panel (g). According to Ref.~\cite{zdanowicz64b}, the \bep\ structure is realized in samples with $\sim\!1.35 \leq x \leq\: \sim\!2.5$. In our own data, tiny indications are already traceable in the XRD data for $x = 1.2$, see e.g., the shoulder formation at the $202_{\beta}$ peak which is split in the case of $x=1.5$. In the data for $x=2.5$ this feature is still clearly discernible but already somewhat weaker. This probably indicates the reappearance of the $\beta$ phase at higher $x$. In panel (g) the positions of selected peaks belonging to the \bep\ structural modification are indicated by red diamond symbols. Based on the data at hand, it is difficult to judge whether the samples with $1.5 \leq x \leq 2.5$ completely crystallize in the \bep\ phase because the two strongest peaks of each structural modification ($202_{\beta}$, $224_{\beta}$ and $204_{\bep}$, $400_{\bep}$) appear at similar $2\theta$ angles.

In summary, we qualitatively confirmed the published structural phase diagram shown in Fig.~\ref{figS1}. The structural transitions from $\alpha$ to $\alpha^{\prime}$ observed in the end compounds \ca\ and \za\  as a function of temperature are also confirmed quantitatively. These produce clear anomalies in specific-heat data (not shown) and are within 3\% (\ca) and 8\% (\za) of the published temperature values. The phase bounderies of the \bep\ dome seem slightly different according to our XRD data but this does not affect the conclusions drawn in this work.

\clearpage
\subsubsection*{\textbf{Lattice parameters}}
\begin{figure}[h!]
\centering
\includegraphics[width=0.8\linewidth,clip]{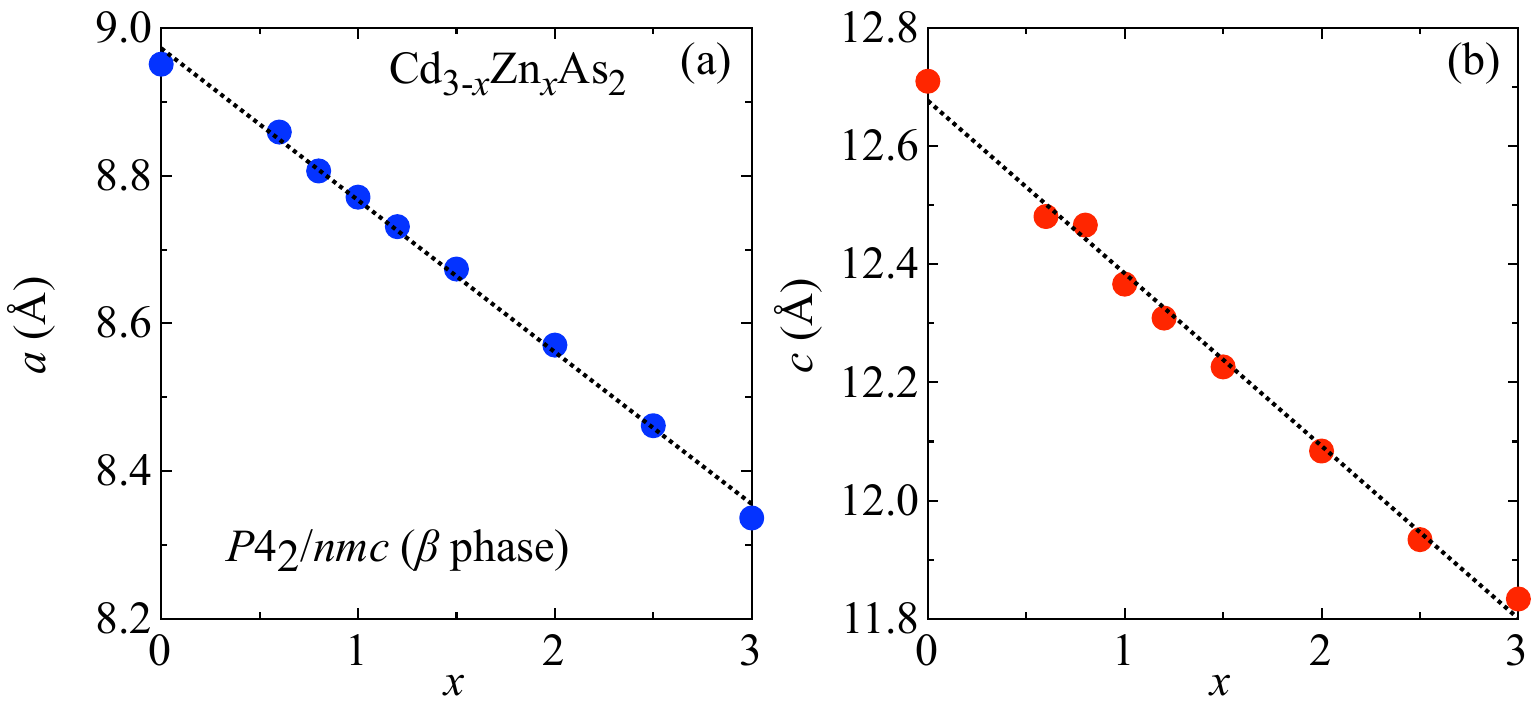}
\caption{Lattice parameters of the (a) $a$ and (b) $c$ axis given in the setting of the $\beta$ structural modification.}
\label{figS4}
\end{figure}
Figure~\ref{figS4} provides the $x$ dependence of the tetragonal lattice constants $a$ and $c$ of \cza\ in the setting of the $\beta$ phase. Both decrease approximately linearly with $x$, satisfying Vegard's law.

\clearpage

\subsection{Results of SEM-EDX analyses}
\label{edx}
\begin{figure}[h]
\centering
\includegraphics[width=1\linewidth,clip]{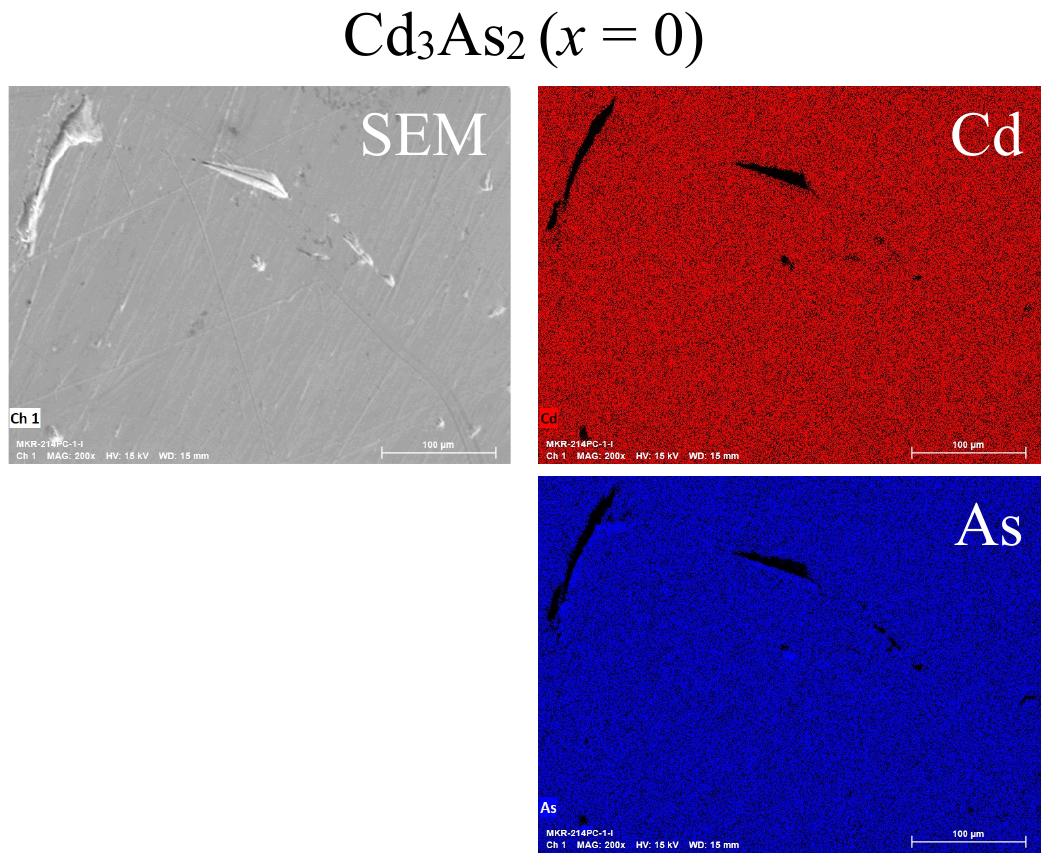}
\caption{SEM-EDX images for \ca\ ($x=0$). The scale bar in all images indicates 100~$\mu$m.}
\label{figS5}
\end{figure}
\begin{figure}[h]
\centering
\includegraphics[width=1\linewidth,clip]{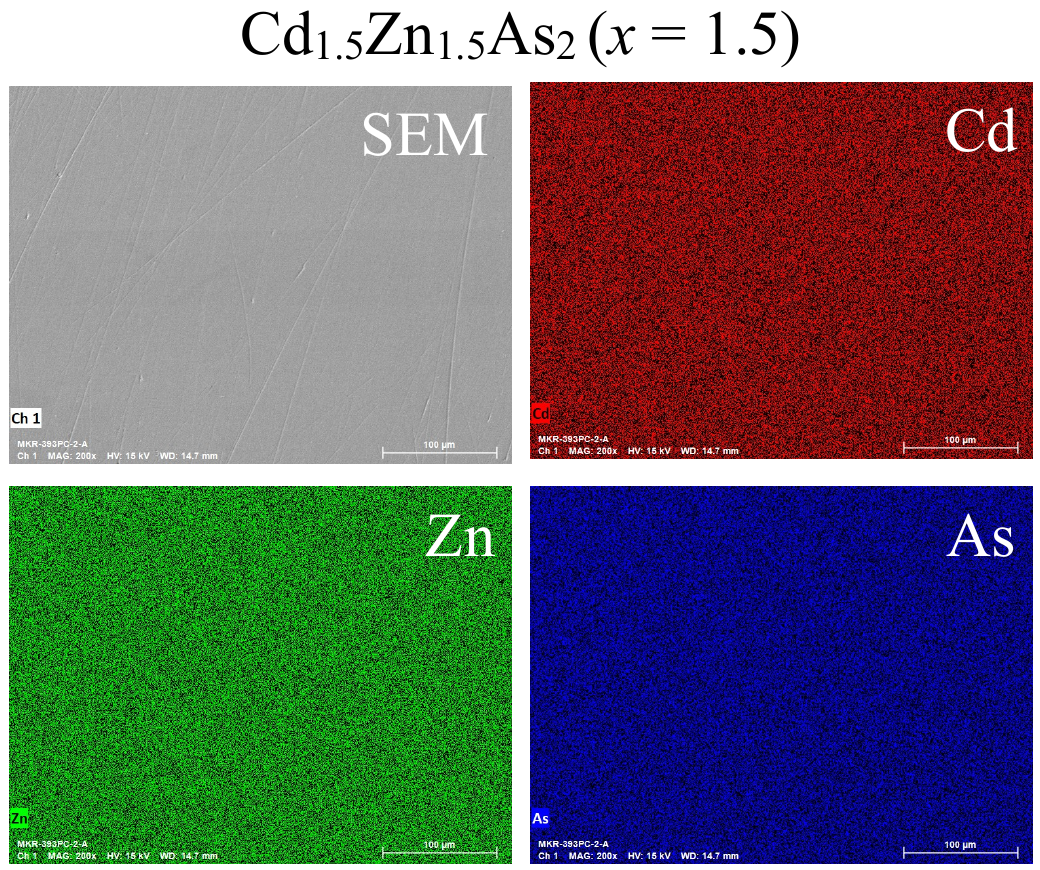}
\caption{SEM-EDX images for Cd$_{1.5}$Zn$_{1.5}$As$_2$ ($x=1.5$). The scale bar in all images indicates 100~$\mu$m.}
\label{figS6}
\end{figure}
\begin{figure}[h]
\centering
\includegraphics[width=1\linewidth,clip]{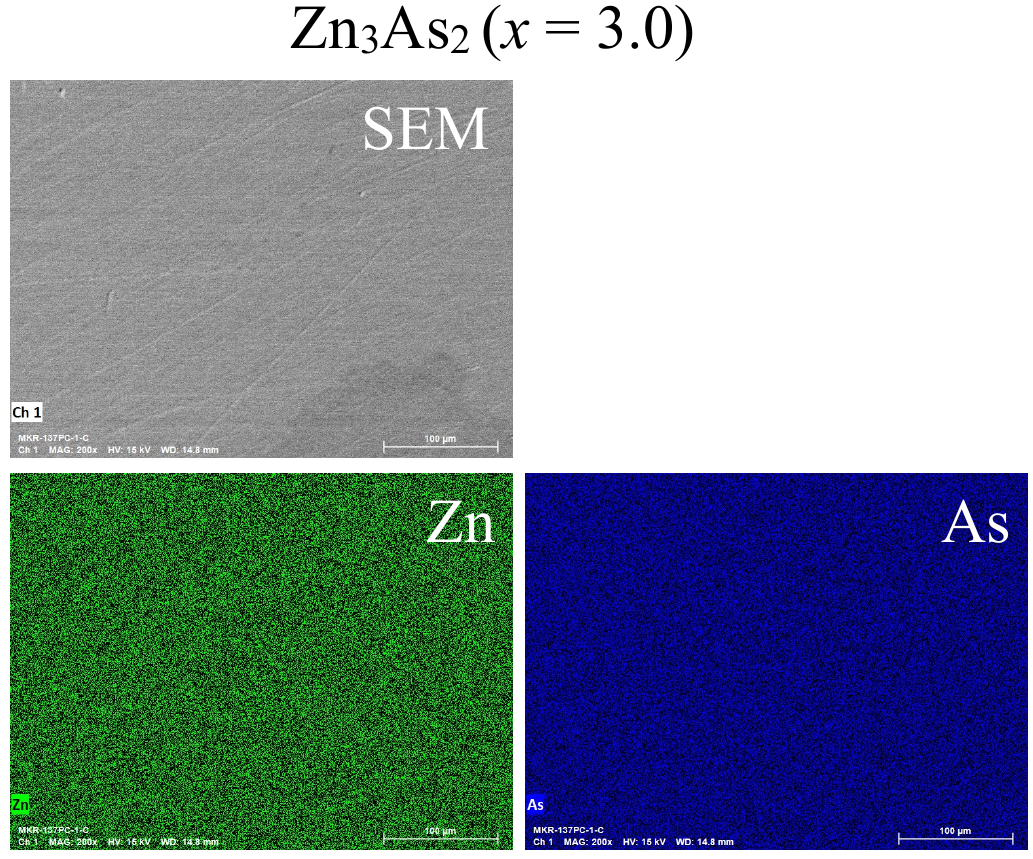}
\caption{SEM-EDX images for \za\ ($x=3.0$). The scale bar in all images indicates 100~$\mu$m.}
\label{figS7}
\end{figure}

\begin{figure}[h]
\centering
\includegraphics[width=1\linewidth,clip]{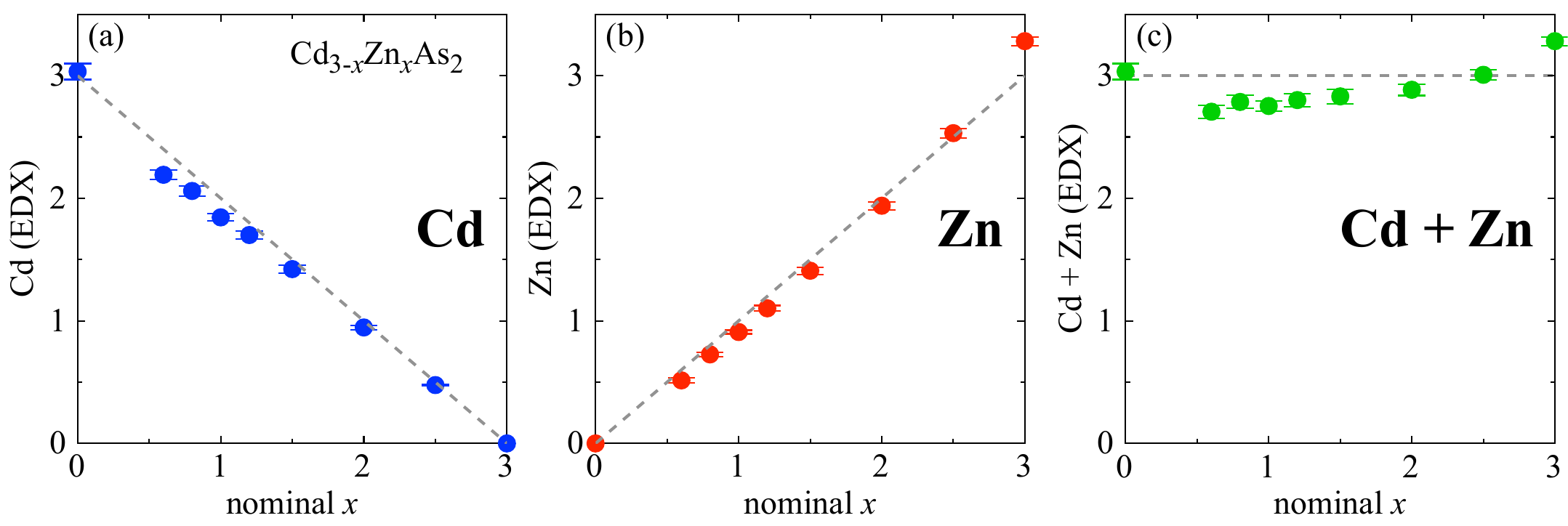}
\caption{Results of EDX analyses of \cza: Estimated chemical compositions of (a) Cd, (b) Zn, and (c) Cd + Zn assuming that the As concentration is 2. The dashed lines in all panels indicate the respective stoichiometric concentration.}
\label{figS8}
\end{figure}

All batches studied were analyzed by SEM-EDX. Selected resulting images are shown in Figs.~\ref{figS5} ($x=0$), \ref{figS6} ($x=1.5$), and \ref{figS7} ($x = 3.0$). These indicate homogeneous distributions of all elements. EDX analyses were carried out on up to 30 spots in different surface areas of a sample. The chemical composition of each spot was estimated assuming that the As concentration is 2. The averaged results are shown in Figure~\ref{figS8} for (a) the Cd concentration, (b) the Zn concentration, and (c) the sum of both. The experimental Cd (Zn) concentration decreases (increases) roughly linearly with the nominal Zn concentration. The sum of the Cd and Zn concentrations slightly exceeds 3 in both pristine materials and falls slightly below in case of the alloyed samples. This deviation from the nominal value is getting smaller with $x$. The error bars are given by the standard deviation calculated from the averaged results of the various EDX analyses and are small throughout the solid solution.
\newpage

\clearpage

\subsection{Details of the thermal conductivity}
\label{kap}
\begin{figure}[b]
\centering
\includegraphics[width=1\linewidth,clip]{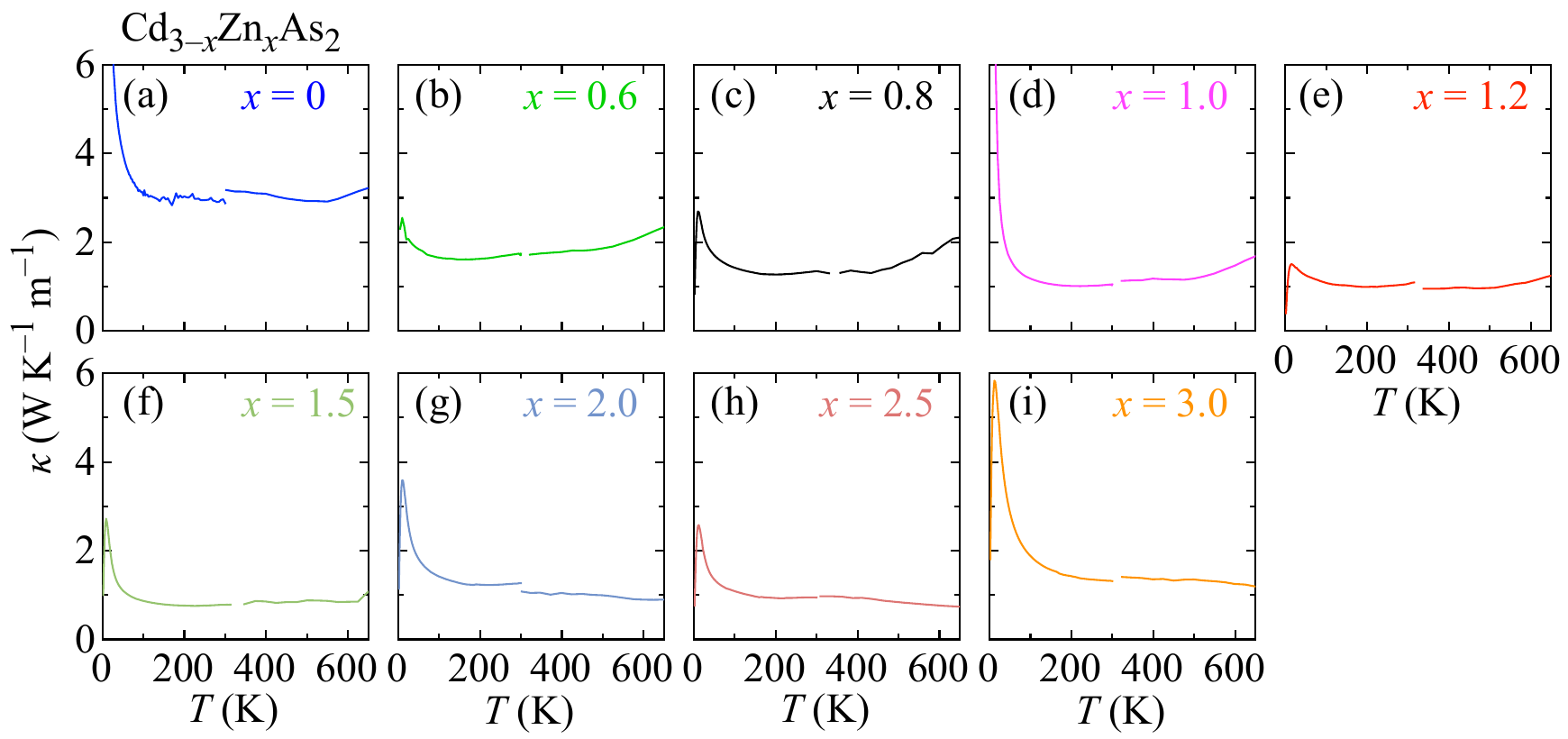}
\caption{Total thermal conductivity in \cza.}
\label{figS9}
\end{figure}

In this Section, the thermal conductivity data for all studied samples are shown in Fig.~\ref{figS9}, including the data which are omitted in Figure~2(d) of the main text. 
As discussed therein, the overall thermal conductivity $\kappa$ is very small and further systematically suppressed as a function of $x$ with a slightly recovered $\kappa$ for $x=3.0$. However, the Umklapp peak at low temperatures is suppressed for the alloyed samples as compared to the pristine compounds except for $x = 1.0$. Conceivable reasons are inhomogeneous samples or differences in the crystal lattice. As argued in Section~\ref{edx}, we can rule out the former. As for the latter, a structural transition below room temperature is unlikely \cite{fujioka21a}. However, the ideal chemical formula of this system is Cd$_4$As$_2$, hence, 25\% of the Cd sites are voids \cite{ali14a}. Upon alloying \ca\ with \za\, Cd, Zn, and the voids have to be arranged on the Cd sites, increasing the disorder. This could cause an additional suppression of the Umklapp peak as compared to the two pristine compounds. For samples in between with an integer amount of Cd and Zn, such as $x = 1$, some additional ordering (``superstructure'') of Cd, Zn, and the voids may be possible, leading to a (partial) recovery of the Umklapp peak. We note that also for $x = 2.0$, a larger Umklapp peak is observed although not as pronounced as for $x = 0$, 1.0, or 3.0.

\begin{figure}[b]
\centering
\includegraphics[width=0.9\linewidth,clip]{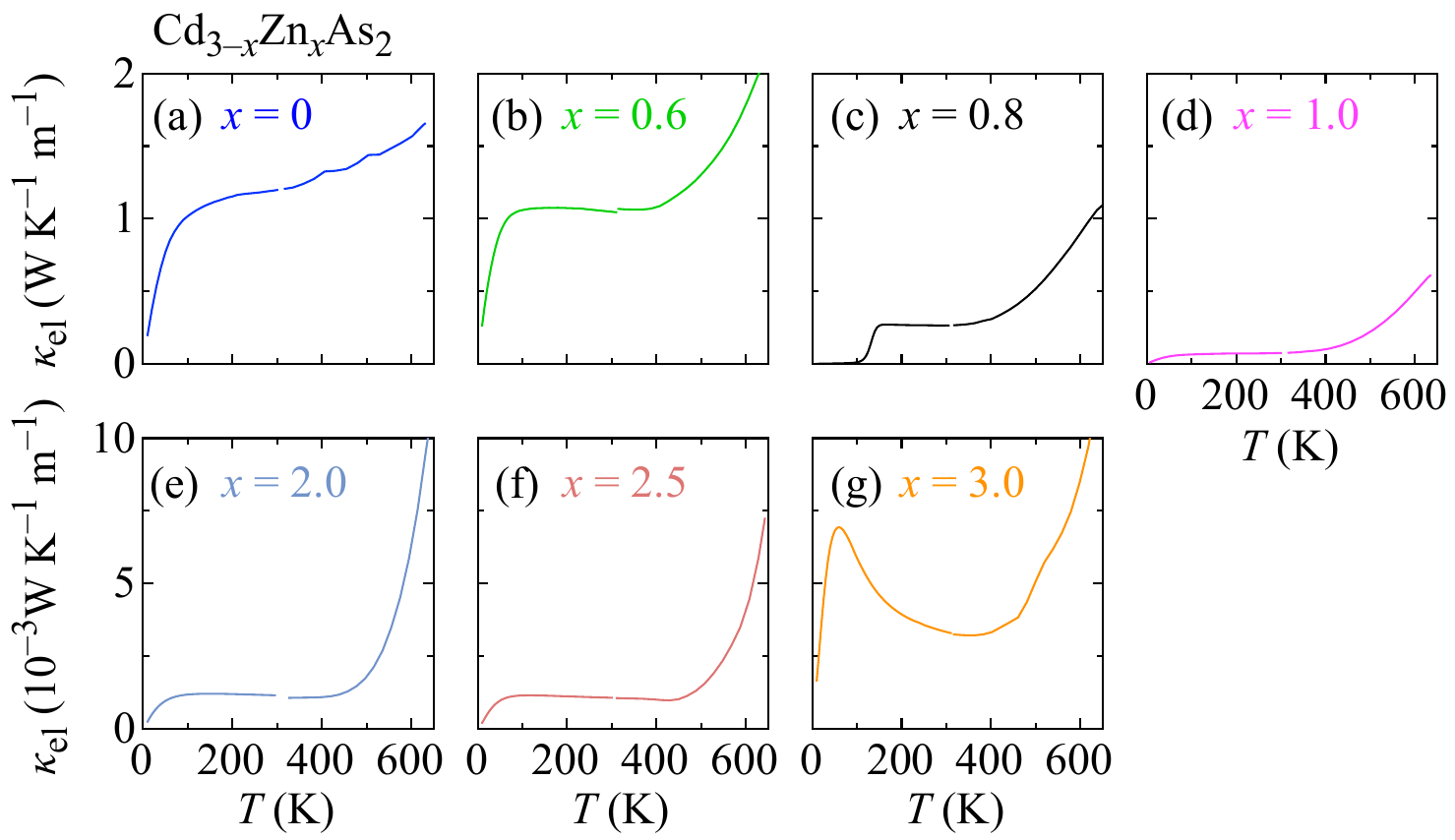}
\caption{Contribution of the charge carriers \kel\ to the total thermal conductivity in \cza. Note the differences in the ordinate scale of panels (a)\,--\,(d) and (e)\,--\,(g).}
\label{figS10}
\end{figure}
\begin{figure}[t]
\centering
\includegraphics[width=0.9\linewidth,clip]{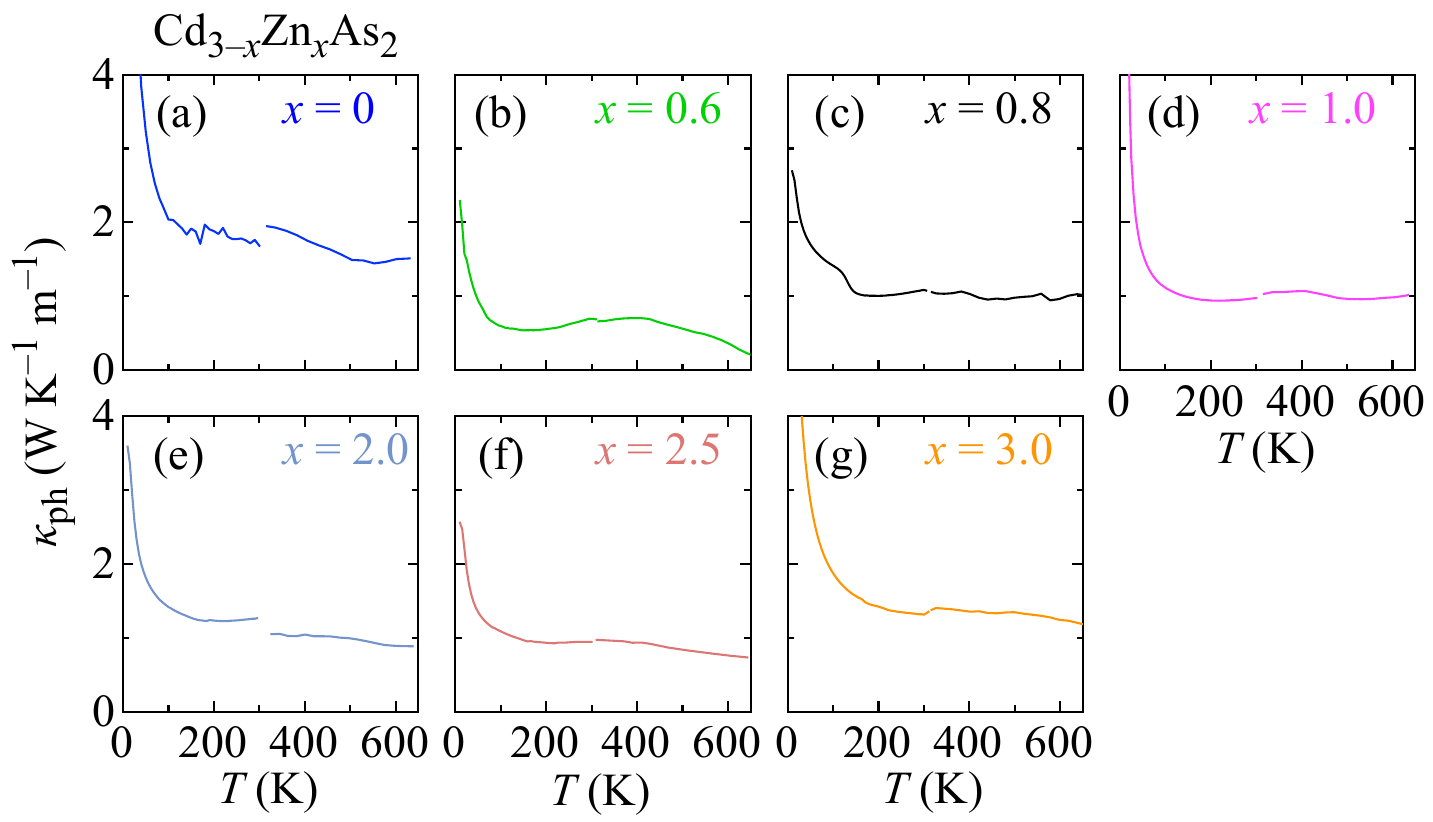}
\caption{Contribution of the phonons \kph\ to the total thermal conductivity in \cza.}
\label{figS11}
\end{figure}
\begin{figure}[h]
\centering
\includegraphics[width=0.8\linewidth,clip]{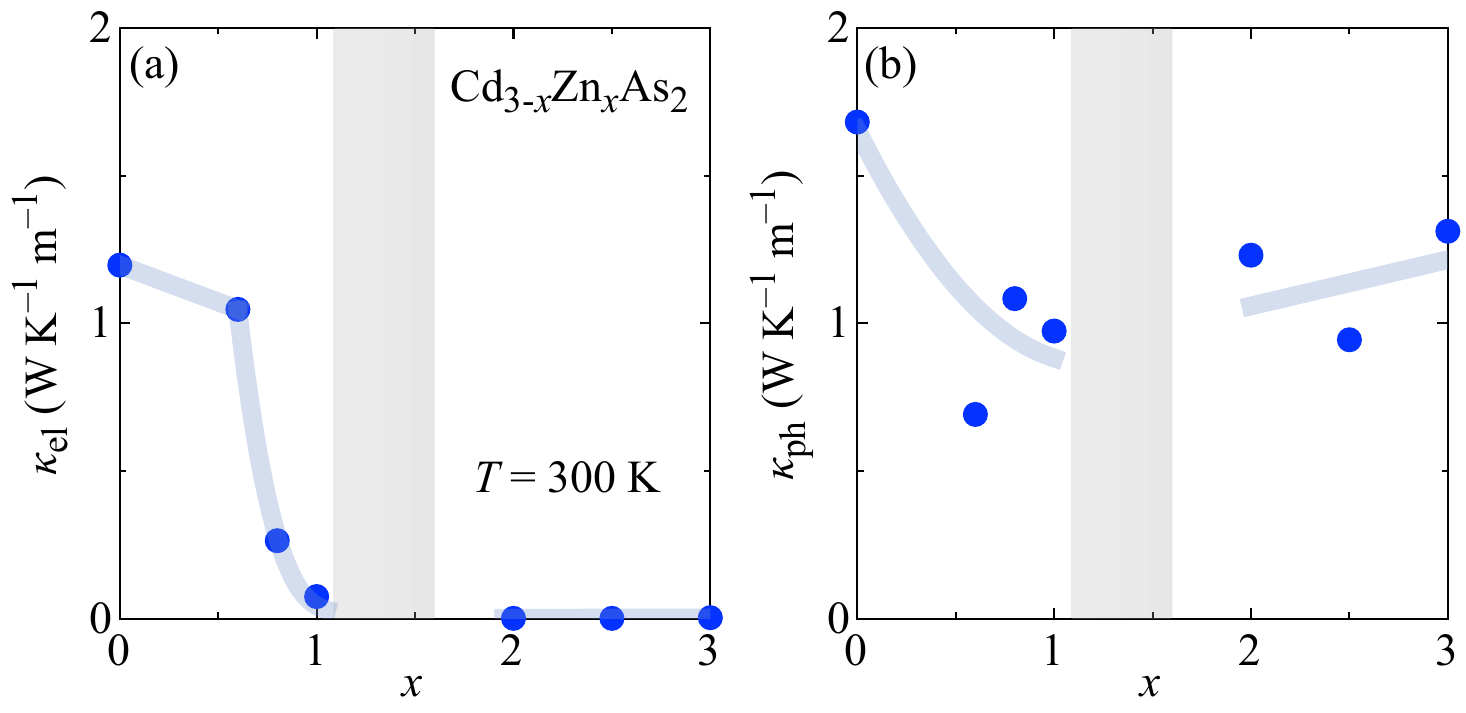}
\caption{Zn concentration dependence of (a) the electronic and (b) the phononic contributions to the total thermal conductivity $\kappa = \kel + \kph$ of \cza\ ($0\leq x \leq 3$) at $T=300$~K. These were obtained by assuming that the Wiedemann-Franz law $\kel=L_0 T/\rho$ with the Lorenz number $L_0 = 2.44 \times 10^{-8}$~V$^2$K$^{-2}$ holds in this solid solution. The grey shaded areas indicate the $x$ range where this law cannot be used due to the simultaneous appearance of electron and hole charge carriers. Respective data points for $x=1.2$ and 1.5 are therefore omitted, see text. In both panels, blueish bold lines are guides to the eyes.}
\label{figS12}
\end{figure}

The total thermal conductivity $\kappa=\kel +\kph$ summarized in Fig.~\ref{figS9} consists of contributions from the charge carriers \kel\ and the lattice \kph. 
Assuming that the Wiedemann-Franz law $\kel=L_0 T/\rho$ with the Lorenz number $L_0 = 2.44 \times 10^{-8}$~V$^2$K$^{-2}$ holds in this system, one can separate them. The resulting breakdown is shown in Figures~\ref{figS10} (\kel) and \ref{figS11} (\kph). Data for $x = 1.2$ and 1.5 are excluded in the latter because in these samples both carrier types, electron and holes, are obviously present and contribute to the electronic and heat transport. In this case, the simple Wiedemann-Franz law does not allow to quantitatively separate these due to contributions from the bipolar thermal conductivity (see, e.g., \cite{may09a}). Respective plots of the $x$ dependencies of \kel\ and \kph\ are shown in Fig.~\ref{figS12}. 

Apparently, the electronic contributions are strongly suppressed with $x$ in agreement with the disappearance of the high-mobility electrons [Figures~3(d) and 3(e) of the main text]. The low-mobility holes which start to appear across $x=1.2$ do not contribute much to $\kappa$. Along with the overall small phononic contributions, the total thermal conductivity is small, which is the main source of the second maximum in $ZT(x)$, as discussed in the main text.

\clearpage
\subsection{Analysis of the Hall resistivity for \boldmath{$x=1.2$} and 1.5}
\label{rhoyx}
\begin{figure}[h]
\centering
\includegraphics[width=0.75\linewidth,clip]{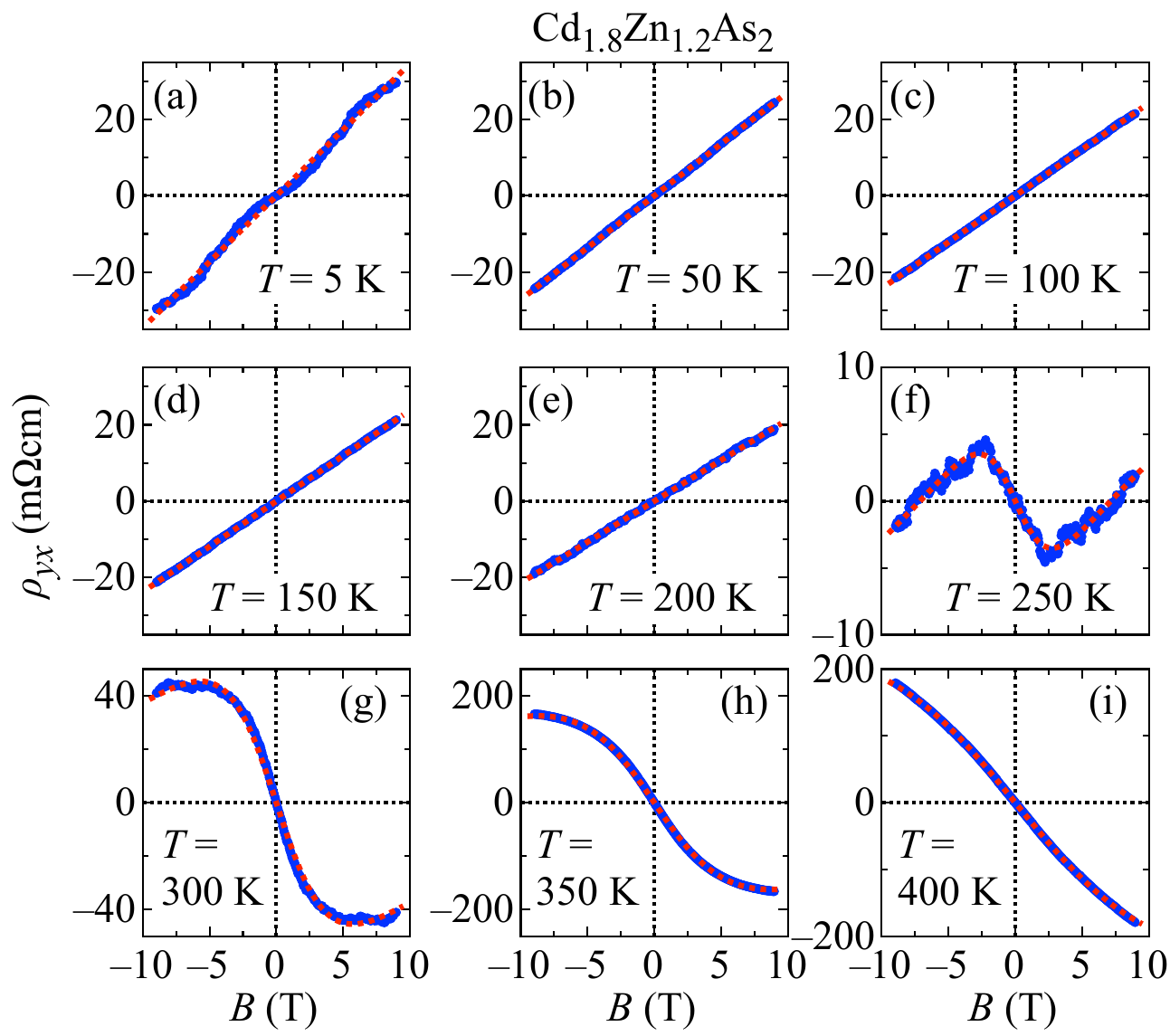}
\caption{Magnetic field $B$-dependent Hall resistivity \ryx\ of Cd$_{1.8}$Zn$_{1.2}$As$_{2}$ at selected temperatures between 5~K and 400~K. Red dotted lines are fits to the data, see text.}
\label{figS13}
\end{figure}
\begin{figure}[h]
\centering
\includegraphics[width=0.75\linewidth,clip]{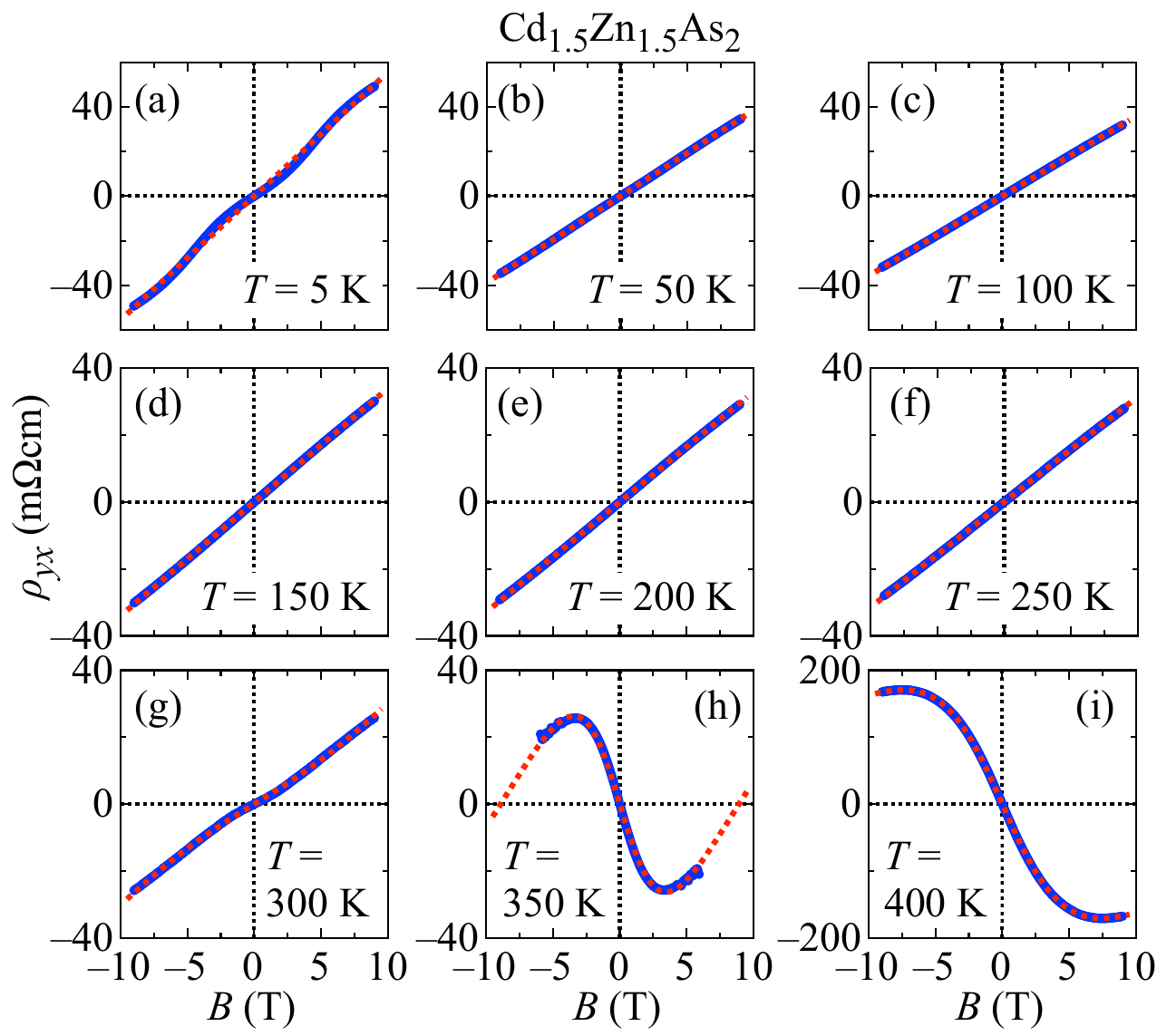}
\caption{Magnetic field $B$-dependent Hall resistivity \ryx\ of Cd$_{1.5}$Zn$_{1.5}$As$_{2}$ at selected temperatures between 5~K and 400~K. Red dotted lines are fits to the data, see text.}
\label{figS14}
\end{figure}
In this subsection, the Hall resistivity and its fitting of Cd$_{1.8}$Zn$_{1.2}$As$_{2}$ (Figure~\ref{figS13}) and Cd$_{1.5}$Zn$_{1.5}$As$_{2}$ (Figure~\ref{figS14}) are discussed. 

For all samples, the Hall resistivity $\ryx(B)$ ($B$ denotes the magnetic field) was measured at selected temperatures between 5~K and 400~K and antisymmetrized with respect to $B$ prior to the analysis. Except for the samples with $x=1.2$ and 1.5, $\ryx(B)$ is either linear with a negative ($x\leq 1.0$) or with a positive slope ($x\geq 2.0$), implying that the charge transport is mainly dominated by one carrier type. The respective charge carrier concentrations were estimated from a simple linear (one-channel) fit. For the two samples with $x=1.2$ and 1.5, the main carrier type changes from electrons to holes as a function of $x$ and temperature. As shown in Figures~\ref{figS13} and \ref{figS14}, at low temperatures, the conduction in both samples is hole-type: $T\leq\: \sim\!200$~K for $x=1.2$ and $T\leq\: \sim\!250$~K for $x=1.5$, respectively. Above, the Hall resistivity is not linear but curved, indicative of two-channel charge transport from electrons and holes. To fit these $\ryx(B)$ data, the two-channel model given in Equation~\ref{Eq1} was employed \cite{maryenko17a}:
\begin{equation}
\label{Eq1}
\ryx(B) = R_{\rm H}\cdot B = \frac{\mu_{\rm e}^2 n_{\rm e}+\mu_{\rm h}^2 n_{\rm h} + (\mu_{\rm e}\mu_{\rm h}B)^2 (n_{\rm e}+n_{\rm h})}
                               {\left((\mu_{\rm e}|n_{\rm e}| + \mu_{\rm h}n_{\rm h})^2 + (\mu_{\rm e}\mu_{\rm h}B)^2 (n_{\rm e}+n_{\rm h})^2\right)e}\cdot B.
\end{equation}
Here, $R_{\rm H}$ denotes the Hall constant, $n_{\rm e}$, $\mu_{\rm e}$ and $n_{\rm h}$, $\mu_{\rm h}$ the charge carrier concentrations $n_{\rm i}$ and mobilities $\mu_{\rm i}$ of the electrons (index e) and holes (index h), and $e= 1.6\times 10^{-19}$~C the elementary charge. In this notation $n_{\rm e}<0$ and $n_{\rm h}>0$. Moreover, one of the four fitting parameters ($\mu_{\rm e}$) was fixed by the simultaneously measured longitudinal resistivity $\rxx(B=0)$ via
\begin{equation}
\label{Eq2}
\rxx(0) = \frac{1}{\left(|n_{\rm e}|\mu_{\rm e} + n_{\rm h}\mu_{\rm h} \right)e}.
\end{equation}
This approach yielded the red dotted fitting curves in Figures~\ref{figS13} and \ref{figS14}. 

\begin{figure}[t]
\centering
\includegraphics[width=0.7\linewidth,clip]{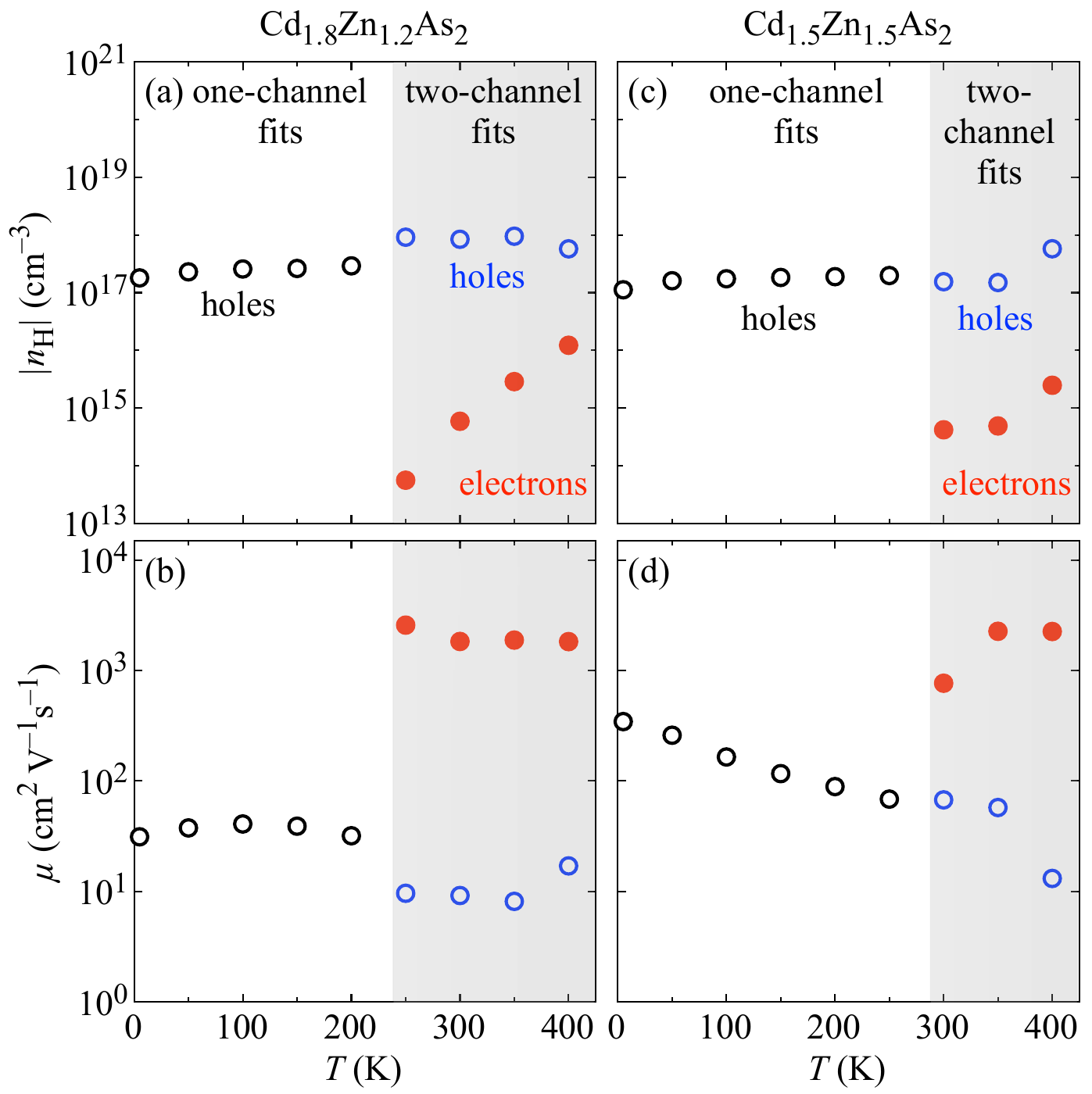}
\caption{Temperature dependence of the (a), (c) charge carrier concentrations and (b), (d) mobilities as estimated from fitting Equation~\ref{Eq1} to the Hall resistivity $\ryx(B)$ of (a) and (b) Cd$_{1.8}$Zn$_{1.2}$As$_{2}$ and (c) and (d) Cd$_{1.5}$Zn$_{1.5}$As$_{2}$. Open symbols refer to hole-type, filled symbols to electron-type conduction. One-channel fit results are given in black, two-channel results in blue (holes) and red (electrons). The gray shaded area in each panel indicates the temperature range in which both carrier types obviously coexist.}
\label{figS15}
\end{figure}

The temperature dependence of the respective charge carrier concentrations and mobilities for both samples are summarized in Figure~\ref{figS15}: The $n_{\rm i}$ are shown in the top panels (a) $x=1.2$ and (c) $x = 1.5$, the $\mu_{\rm i}$ in the bottom panels (b) $x=1.2$ and (d) $x=1.5$, respectively.
In both samples the electron concentration increases with temperature but on a very low level $<\: \sim\! 10^{16}$~cm$^{-3}$, i.e., almost depleted while they keep their high mobility $>\: \sim\! 10^3$~cm$^{2}$V$^{-1}$s$^{-2}$. Holes dominate the charge transport with a concentration of $10^{17}$\,--\,$10^{18}$~cm$^{-3}$ but exhibit a much smaller mobility $<\: \sim\! 10^2$~cm$^{2}$V$^{-1}$s$^{-2}$. 

\clearpage
\subsection{Analysis of the longitudinal resistivity at high temperatures}
\label{arrh}
\begin{figure}[h]
\centering
\includegraphics[width=1\linewidth,clip]{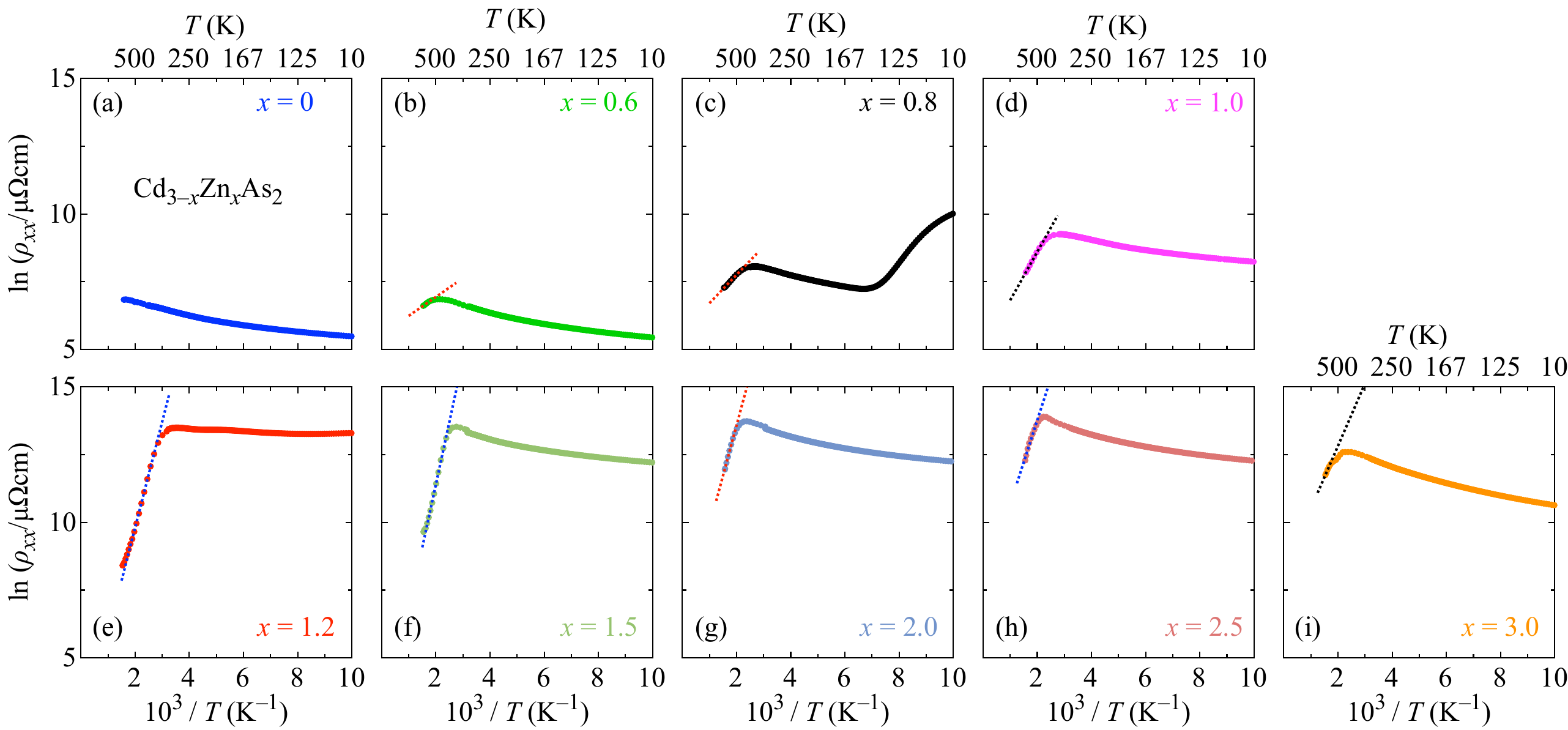}
\caption{Longitudinal resistivity of \cza\ displayed as Arrhenius plots $\ln \rxx$ vs $T^{-1}$. The top axes give the temperature in K. The dotted lines in panels (b)\,--\,(i) are linear fits to the high-temperature part of the resistivity.}
\label{figS16}
\end{figure}
Figure~\ref{figS16} displays the longitudinal resistivity as Arrhenius plots $\ln \rxx$ vs $T^{-1}$ along with linear fits (dotted lines in the panels for $x\geq 0.6$) to the high-temperature part of \rxx\ above the maximum in the data shown in Figure~2(a) of the main text. From the slopes of the linear fits, the respective activation energies $\Delta_{\rm Arr}(x)$ were estimated. These are discussed along with a comparison with other characteristic properties in the next Section. 
\clearpage
\subsection{Comparison of additional properties}
\label{comp}
\begin{figure}[h]
\centering
\includegraphics[width=0.8\linewidth,clip]{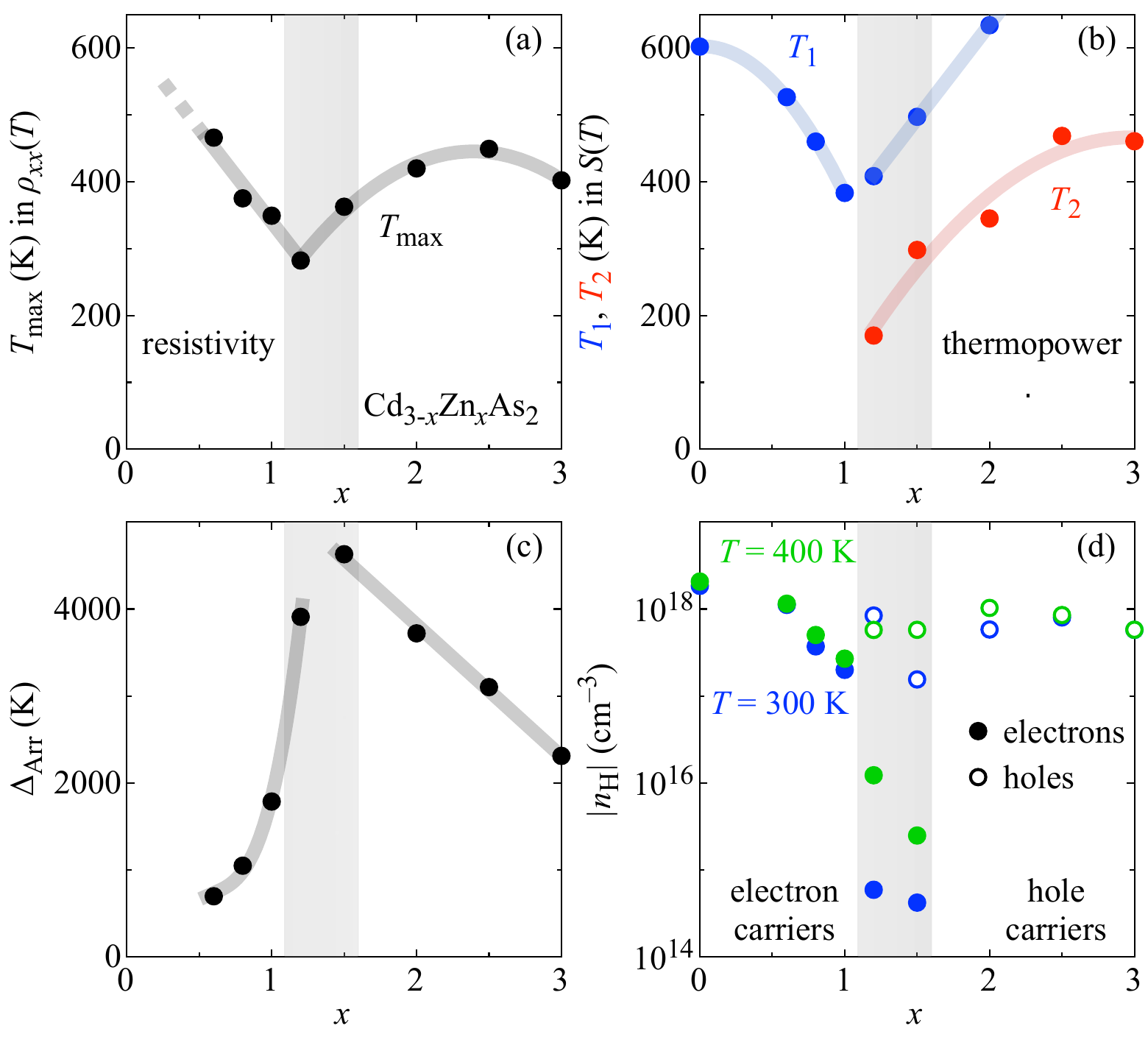}
\caption{Zn concentration dependence of (a) the temperature \Tmax\ of the maxima in resistivity data, (b) the temperatures \Tmin\ (blue) and \TSmax\ (red) of the extrema in thermopower data, (c) the activation energy \DAr\ as estimated from the linear fits shown in Figure~\ref{figS16}, and (d) the absolute value of the charge carrier concentration $|\nH|$ estimated from Hall-resistivity data at 300~K (blue) and 400~K (green). In the latter, filled and open symbols indicate electron- and hole-type conduction, respectively. The gray shaded area in each panel indicates the crossover range $\sim\! 1.2 \leq x \leq\: \sim\! 1.5$ where the dominant carrier type changes from electrons to holes as discernible in (d). The bold lines in panels (a)\,--\,(c) are guide to the eyes.}
\label{figS17}
\end{figure}
\clearpage

Figure~\ref{figS17} summarizes the $x$ dependence of several characteristic features observed in \cza. Panel (a) shows the $x$ dependence of the extrema \Tmax\ in the temperature dependence of the resistivity [Figure~2(a) of the main text], which decrease linearly with $x$ up to $x\sim 1.2$. Above, $\Tmax(x)$ starts to increase and peaks around $x\sim 2.5$. 

Figure~\ref{figS17}(b) shows the respective extrema \Tmin\ and \TSmax\ observed in the temperature dependence of the thermopower data [Figure~2(b) of the main text]: The extrema \Tmin\ decrease up to $x = 1.0$ and increase above. For $x\geq 1.2$, $\TSmax(x)$ in $S(T)$ evolves qualitatively similar as \Tmax\ in $\rho(T)$. However, the extrema \TSmax\ are missing in $S(T)$ for $x\leq 1.0$. 

The $x$ dependence of the activation energy \DAr\ as estimated from the Arrhenius plots shown in Figure~\ref{figS16} is plotted in Figure~\ref{figS17}(c). It exhibits an increase with $x$ for $x\leq 1.2$. By contrast, for $x\geq 1.5$ it decreases linearly with $x$. 

Figure~\ref{figS17}(d) summarizes the $x$ dependence of the absolute charge carrier concentration $|\nH|$ as estimated from Hall-resistivity measurements at 300~K and 400~K, with the latter being the highest temperature at which we can measure $\ryx(B)$. The 300-K data are replotted from Figure~3(d) of the main text. Comparing these data, for most samples the charge carrier concentration slightly increases as a function of temperature.

As for the situation below $x\sim 1.2$, the existence of maxima in $\rho(T)$ indicates that there are additional thermally activated charge carriers at elevated temperatures. These are reflected in the increase in \nH\ at a given $x$ when going from 300~K to 400~K. It is probably safe to assume that this tendency holds up to higher temperatures, given the overall behavior of the resistivity shown in Figures~2(a) and 3(a) in the main text. Together with the decrease in $|\nH(x)|$, this suggests that the Fermi level shifts downwards with $x$ up to at least $x \sim 1.0$. As seen in the band structure plot in Figure~1(a) of the main text, there are indeed additional bands within a few 100~meV of \EF, to which carriers can be excited in qualitative agreement with this picture. However, this scenario cannot explain the behavior observed upon further increasing the Zn concentration: the $x$ dependencies of the properties shown in Figures~\ref{figS17}(a)\,--\,(d) change drastically, pointing toward a significant modification of the band structure. 

Altogether, these findings further support the conclusion drawn in the main text that the band structure up to $x\sim 1.0$ is of the \ca\ type, i.e., topologically nontrivial, and that this changes across $x \sim 1.2$, beyond which the system is dominated by the topologically trivial band structure of the \za\ type. We note that even in nominally pure \za, the Fermi level does not lie inside the band gap but in the valence bands due to unintentionally doped holes. Upon increasing $x$ above 1.2, \EF\ shifts slightly downwards toward the end compound \za\ as indicated by the increase in the hole concentration. 

\clearpage
\subsection{Sample dependence of transport data in \cza}
\label{reprod}
As for the comparison with our previous publication \cite{fujioka21a}, the present results are qualitatively in good agreement. However, quantitatively this is not always the case in resistivity and thermopower measurements as, e.g., manifested in the occasional observation of a metal-insulator transition below room temperature in samples with $x\leq 1.2$, cf.\ Ref.~\cite{fujioka21a}. The origin seems twofold: 

(i) We observe a strong sample dependence even among samples cut from the same batch, which had been also noticed in the literature, see, e.g., Refs.~\cite{ito77a,tliang14a,crassee18a}. A possible explanation can be given when taking into account that there is an intrinsically large number of defects in \ca\ (and \cza). As argued in Section~\ref{kap}, chemically the stoichiometric formula of this system should be Cd$_4$As$_2$ and, hence, 25\% of the Cd sites are empty \cite{zwang13a,ali14a}. It was proposed that the ordering of these voids in \ca\ can differ from sample to sample \cite{ali14a,tliang14a,prvComm} causing the differences. From our own experience with this system, we can add here that the results of resistivity and thermopower measurements can vary quantitatively even when measuring on the same sample surface, e.g., when the electric contacts were made on different positions on the same surface. This effect seems to become even worse when introducing Zn on the Cd site. 

(ii) We also observe an annealing effect above room temperature, which tends to affect the resistivity and the thermopower and modify their temperature dependencies. This explains the slightly smaller room-temperature values found here for $ZT$ as compared to our previous work, which was restricted to $T<300$~K.


%